\documentclass[12pt]{aastex62}

\begin{document}

\parindent=1.0cm

\title{The Digitization of Photographic Spectra in the Dominion Astrophysical 
Observatory Plate Collection with Commercial Scanners: A Pilot Study} 

\author{T. J. Davidge}

\affiliation{Dominion Astrophysical Observatory,
\\Herzberg Astronomy \& Astrophysics Research Center,
\\National Research Council of Canada, 5071 West Saanich Road,
\\Victoria, BC Canada V9E 2E7\\tim.davidge@nrc.ca; tdavidge1450@gmail.com}

\begin{abstract}

	Commercial flatbed scanners have the potential to deliver a 
quick and efficient means of capturing the scientific content of spectra 
recorded on photographic plates. We discuss the 
digitization of selected spectra in the Dominion Astrophysical Observatory 
(DAO) photographic plate collection with commercial scanners. 
In this pilot study, emphasis is placed on assessing if the 
information on the plates can be recovered using Epson V800 and 
12000XL scanners; the more complicated issues associated with the 
shortcomings of photographic materials, such as correcting for nonlinearity, 
is deferred to a future study. Spectra of Vega ($\alpha$ Lyr) that 
were recorded over $\sim 4$ decades with the DAO 1.8 meter 
telescope are examined. These spectra sample a range of photographic 
emulsions, plate preparation techniques, calibration 
information, observing techniques, and spectrograph configuration. 
A scanning density of 2400 elements per inch 
recovers information in the spectra. Differences in the 
modulation transfer function (MTF) of the two scanners are found, with the 
Epson 12000XL having a superior MTF. Comparisons with a CCD spectrum of Vega 
confirm that moderately weak features are faithfully recovered in 
photographic spectra that have been digitized with the 12000XL scanner. The 
importance of scanning the full plate to cover the light profile of the 
target and calibration information is emphasized. Lessons learned from 
these experiments are also presented. 

\end{abstract}

\section{INTRODUCTION}

	The Dominion Astrophysical Observatory (DAO) hosts an extensive 
collection of photographic spectra that were recorded 
throughout much of the last century. The collection includes spectra 
that were obtained with the DAO 1.8 meter telescope (commissioned 1918), 
the DAO 1.2 meter telescope (commissioned 1962), and the 0.4 meter 
telescope at the Dominion Observatory (DO) in Ottawa (commissioned 
1905). The spectra recorded at the latter facility date back to 
1915, and so are the oldest in the collection.

	The DAO collection contains spectra that were recorded over many 
decades, and so digitized versions of the plates are a 
potentially important resource for long-term studies of 
stellar properties. However, there are various complications 
associated with the digitization of these spectra. 
From a purely pragmatic perspective, digitizing the tens of thousands of 
plates in the DAO collection in an efficient manner that captures 
their full information content, including all calibration 
information, presents a formidable logistical challenge. 
Another challenge is the construction of an archive of digitized material that 
links all available ancillary material for each plate. 
Yet another challenge is providing data products 
given that the plates contain a diverse range of calibration information 
for calibrating the wavelength scale and assessing 
nonlinearity. An unfortunate related problem is that many members of the 
community who are familiar with photographic materials and 
their limitations are either aged or deceased. This paper deals with 
the digitization of the spectra, as the capture of information in a digital 
format is a fundamental first step for enabling further examination of 
these data.

	An instrument that has been commonly employed in the past to digitize 
photographic spectra is the Photometric Data Systems (PDS) microdensitometer. 
While providing detailed control over parameters that are important for 
recovering scientific content, digitization with a PDS can be a slow process, 
and this has spurred the development of customized devices for 
the digitization of large format plates \citep[e.g.][]{sto1984}. These devices 
typically require specialized environments, and so are not portable.
Challenges that arise due to vibration, thermal 
stability, and warm-up time have been discussed by \citet{teu1984}.
Still, efforts to digitize moderately large ensembles of photographic 
material using microdensitometers have been undertaken; arguably one 
of the most ambitious such endeavours was the scanning of the Palomar 
Observatory Sky Survey and the UK Schmidt survey to construct products such 
as the Guide Star Catalogue \citep[e.g.][]{lasetal1990}. 

	There are many tens of thousands of plates in the DAO collection, 
and so it is important to identify an efficient means of scanning 
this material and recovering the information that they 
contain. This can not be accomplished simply by 
restricting the area on each plate to be scanned, as a 
typical photographic spectrum contains useful information 
over a large part of its area. For example, the seeing profile 
broadens the spectrum perpendicular to the dispersion direction, 
and the profile should be sampled in its entirety to enhance 
the signal-to-noise (S/N) ratio. There are calibration regions that can be 
used to determine the characteristic curve of the plate. 
The seemingly empty areas of each plate also contain information 
for assessing scattered light, fogging, and 
blemishes (e.g. fingerprints from incorrect handling of the unexposed 
plates, hair, dust etc); corrections for these 
are needed if the ultimate goal is to produce a digitized spectrum that 
can approach the quality of those recorded with a CCD. When considered 
together, these factors make scanning a large collection of photographic 
spectra with a device such as a PDS a daunting task.

	The current generation of commercial scanners offer an intriguing 
alternative to PDSs. The potential benefits of scanners are 
obvious; they are easy to use and can be deployed on an office 
desk. They do not require specialized set-up and 
training for operation. Moreover, an entire plate is scanned in 
one pass, allowing the complete slit profile and the full range of calibration 
information to be sampled. Still, the ability of earlier generations of these 
devices to deliver science-caliber data has been questioned, due primarily to 
the limited resolution of the optics \citep[e.g.][]{sim2009}. 
However, there have since been successful applications 
in which recent models have been employed to 
digitize images \citep[]{yuletal2019, ceretal2021, gluetal2022}. 

	In this study we adopt an empirical approach to 
assess if the current generation of desktop scanners 
can capture the core science content of photographic spectra. 
This is done by scanning a modest set of spectra with two off-the-shelf devices.
The ultimate benchmark for assessing the results is a comparison with spectra 
recorded with a CCD detector. Such a comparison necessitates the use 
of a star with stable spectroscopic properties, and Vega is selected for the 
current work.

	Vega has a long history of use as a reference star. It 
has stable H$\alpha$ absorption \citep[]{chaandmey1985}, although 
\citet{bohetal2012} conclude that it is a `quiet' pulsator. 
\citet{gra2007} and \citet{hiletal2010} discuss other issues 
that may affect its status as a standard star, although these will 
not affect the conclusions reached in this study. In addition to
spectrophotometric stability, the spectrum of Vega 
contains a rich collection of features that are useful 
diagnostics for assessing the recovery of science content. The deep Balmer 
absorption lines in the spectra of Vega and other early A stars are important 
for assessing linearity and wavelength resolution, as these lines sample a 
broad range of intensities along the characteristic curve. 
Weak metallic absorption features in the Vega spectrum
allow additional limits to be placed on scientific utility and linearity.

	There are also purely pragmatic reasons for selecting Vega as a 
reference. It is one of the brightest stars 
in the northern sky, and it transits at midnight during the 
summer when the skies above the DAO tend to be clear. 
There are thus numerous high-quality photographic spectra of this star in the 
plate collection that span the entire period of time that 
plates were in use. There are also a number of spectra of Vega recorded with 
a CCD detector, and comparisons with these observations provide an 
important means of gauging the information contained in a digitized spectrum. 

	The scanning experiments discussed in this paper 
are restricted to spectra recorded with the 1.8 metre telescope, which 
constitute the largest part of the DAO plate collection. These spectra 
span a multi-decade timeline, and were recorded with a range of observing 
conditions, instrumental configurations, and plate preparation techniques. The 
oldest of the spectra have minimal or no linearity calibration information, but 
almost all have on-plate arc spectra for wavelength calibration.
Many of the spectra were recorded for radial velocity studies, with emphasis 
on measuring the centroids of prominent lines, as opposed to quantitative 
measurements of line shape and strength. The influence of the radial velocity 
emphasis when extracting spectra for other purposes is discussed later in 
the paper.

	There are a number of issues to be considered when dealing with 
spectra retrieved from photographic plates with the ultimate goal of creating 
a scientifically useful digital archive. Paramount among these is the 
identification of procedures that will allow the information on a plate to 
be captured in a digital format. The current paper focuses on this issue by 
examining a small number of photographic spectra with the goal 
of assessing the viability of digitization with a commercial scanner. We 
emphasize that this paper does not deal with other 
issues such as capturing metadata and the 
development of software that is required to extract detailed 
scientific information from the spectra, although 
calibration information is examined to estimate signal levels where 
the response is expected to be linear. As demonstrated later in this paper, 
the results are promising, thereby encouraging future digitization 
efforts with commercial scanners. 

\section{PLATE SELECTION}

	Glass is a fragile storage medium, requiring 
protection from hazards that might cause it to break or crack. 
The plates at DAO are stored in individual envelopes 
that are placed upright in metal filing cabinets. 
The cabinets are in a temperature-controlled environment, and are stored 
with a complete collection of observing logs. Envelopes are labeled with 
the target name, location on the sky, the date of acquisition, 
and other information, the nature of which depends mainly on when the data 
were recorded. There are an estimated $60000+$ spectra in the 1.8 metre plate 
collection. While not the subject of this work, there are also a large number 
of direct imaging plates that were recorded at the telescope's Newtonian focus.

	The main criteria for selecting plates to be used in this study 
were (1) a signal level that produced a visually prominent spectrum in 
which at least the deepest absorption lines are 
detected, and (2) a `good' cosmetic quality, such that 
the areas of the plates that contain the stellar spectra and 
calibration information are not damaged or cracked. The assessment 
of signal level was done visually by searching for clear detections 
of Balmer lines. Selection with this criterion is a straightforward 
task given the depths of these lines in the Vega spectrum, coupled with the 
brightness of Vega, which means that there are few spectra of this source 
in the DAO collection in which lines are not detected. 
There are a large number of Vega spectra in the DAO collection, 
and no effort was made to identify the 'best' spectrum at a given epoch. 
A cursory visual examination, such as that employed here, is the 
sort of procedure that a user might employ when conducting an initial 
search for spectra of potential interest in an undigitized plate collection. 

	The intent of the first criterion is to cull noisey spectra, 
and it also biases the sample toward spectra in which much of the 
signal samples the middle and upper portions of the characteristic curve. 
As demonstrated later in this paper, there is a 
compelling reason to select spectra where strong lines can be 
detected, as these are helpful to examine characteristics 
such as nonlinearity and the overall scientific 
utilty of a scanned spectrum. As for the second criterion, an effort has 
been made to identify plates with cosmetics that are typical of the 
collection in general. Plates that tend to show evidence of low-level 
fogging and/or modest cosmetic defects such as scratches and/or 
detritus embedded in the emulsion are the norm. All of the plates in the 
DAO collection contain writing that includes the plate 
number and source name. None of these artifacts overlapped with 
the spectrum and calibration information. 

	A final criterion was the central wavelength of 
the observations, which is important to enable plate-to-plate comparisons. 
A large fraction of the spectra recorded with the 1.8 metre 
telescope sample wavelengths between 0.38 and $0.48\mu$m. This wavelength 
interval was intentionally targeted in the early years of 1.8 meter 
operation to accommodate the radial velocity studies that were the 
motivation for building the telescope. This wavelength interval also contains 
a rich population of Balmer lines and numerous atomic transitions from metals 
in the Vega spectrum. Hence, spectra that are approximately centered near 
H$\gamma$ were selected for this work.

	The selection process described above is admittedly 
qualitative. However, the goal was not to identify the best spectra 
of Vega, but those that appear to be among the best at a given epoch in 
terms of line detection. In fact, later in the paper it is 
shown that the spectra sample a range of S/Ns and 
line depths, as is to be expected based on factors such as emulsion type, 
hyper-sensitivation methods (if any), and instrument 
throughput. Future users of undigitized plate archives may 
impose stricter quantitative requirements on spectra selected for use, based 
on the intended scientific purpose. In this regard, users might consider 
conducting an initial assessment of plates based on a significantly lower 
spatial digitization frequency and digitization format (e.g. 8 bit sampling) 
than that ultimately adopted for full scientific 
use, as spectra obtained with even comparatively low density scans provide 
hints to scientific potential (e.g. Figure 5). Such a time-saving approach 
to conduct an initial assessment of spectroscopic content might 
prove to be of great use when examining very large format plates, such as 
grism or grens surveys conducted at prime focus.

	Plates containing spectra of Vega that were recorded 
over a four decade timeline were selected for digitization. 
These sample different emulsions, instrumentation, and plate preparation 
procedures. Details of the plates, including information taken from the 
nightly observing logs, are provided in Table 1. 
There are multiple spectra of Vega on the 1949 and 1963 plates, and these 
allow an examination of intraplate homogeneity and linearity 
that should be independent of emulsion type and hypersensitization procedure. 

\begin{center}
\begin{deluxetable}{lcccl}
\tablecaption{Plates used in this Study}
\tablehead{Date & Plate \# \tablenotemark{a} & Seeing\tablenotemark{b} & Plate Type/\tablenotemark{b} & Notes \\ & & (arcsec) & Emulsion & }
\startdata
August 14, 1925 & 12101 & 1.5 & E-40 & Minimal linearity information\\
June 27, 1937 & 26881 & 4 & E-40 & No linearity calibration. \\
 & & & & Recorded in twilight. \\
October 20, 1949 & 40452 & 4 & Cr H.S.\tablenotemark{c} & 3 spectra - 30/15/10 seconds; No arcs \\
July 5, 1963 & 59216 & 4 & IV-0 & 4 spectra \\
\enddata
\tablenotetext{a}{Internal numbering scheme for 1.8 meter plates.}
\tablenotetext{b}{Information taken from the observing logs.}
\tablenotetext{c}{H.S. = hypersensitized}
\end{deluxetable}
\end{center}

	The design of the original spectrograph is described by 
\citet{pla1918,pla1924}, and the instrument described in those papers 
was used to record the 1925 and 1937 plates. 
Major changes were made to the spectrograph in the mid 1940s to 
deliver higher spectral resolutions \citep[]{beaetal1946}, and the 
1949 and 1963 spectra were recorded with this configuration. While provisions 
were made to use gratings as part of this upgrade, 
prisms were used as dispersive elements up until at least the mid 1950s 
\citep[e.g.]{ricandmck1956}. Informal discussions with staff suggest 
that minor modifications were made to the instrument throughout the 
time span of the observations discussed here, but these were done 
without archived documentation.

\section{SCANNING}

	The spectra were digitized with Epson V800 
and 12000XL scanners. Information about 
both devices can be found on the Epson corporate website 
\footnote[1]{https://files.support.epson.com/docid/cpd4/cpd41530.pdf and 
https://files.support.epson.com/docid/cpd5/cpd53120.pdf}. 
The V800 is compact, light weight, relatively 
inexpensive, and is intended for home and office use. It 
has a $216 \times 297$ mm scanning surface, and so 
multiple 1.8 meter photographic spectra can be scanned at once.
A maximum resolution of 4800 dots per inch (DPI) over the entire scanner 
surface is advertised, with even higher scanning densities 
claimed over smaller areas. Still, \citet{sim2009} 
found that the modulation transfer function (MTF) of a similar 
Epson scanner (model V750) did not allow the advertised maximum 
resolution to be realized in practice. Evidence is presented later in 
this paper that the MTF of the V800 scanner produces fuzzier images 
than the 12000XL scanner when scanning at 2400 dpi. 

	When it was purchased by DAO in late 2022, the Epson 12000XL was the 
premier Epson scanner for high-quality artistic and commercial use. 
A working assumption for purchasing this device was that the MTF would 
be superior to that of its smaller, less expensive V800 cousin. The 12000XL 
has a $310 \times 437$ mm scanning surface, and so there is a multiplex 
advantage when compared with the V800 scanner. Nevertheless, there is a 
trade-off between scanning density and area coverage, 
as internal memory limitations mean that full spatial coverage 
can not be realised when scanning at the highest sampling densities. 
While resolutions as high as 2400 dpi are advertised over the entire 12000XL 
scanning surface, this was found to hold only for lower contrast scanning 
modes, and so is not appropriate for the scanning of spectra. A maximum 
resolution of 4800 dpi is advertised for smaller areas, although such 
high resolutions may not be needed for photographic plates given the 
$\sim 10 - 30\mu$m resolution of photographic emulsions. 
A 12000XL scanner has been used to digitize 
photographic images in the Yerkes Observatory plate collection 
\citep[]{ceretal2021, gluetal2022}. \citet{ceretal2021} 
scanned a r'{e}seau screen with a 12000XL scanner, and conclude that 
features that are `significantly smaller than $50\mu$m' can be resolved. 
For comparison, the 2400 dpi scanning density that is adopted for the current 
work corresponds to $\sim 10\mu$m sampling.

	All plates were scanned with 16 bit sampling, 
and some were also recorded with 8 bit sampling. 
While 8 bit sampling obviously compromises contrast, it was considered 
as a possible initial diagnostic to assess 
the quality of spectra in an undigitized archive. The final spectra 
obtained with 8 bit sampling are similar to those recorded with 16 bit 
sampling, likely because spectra are extracted over a range of locations along 
the slit (see next section). When information from the different locations 
is combined then digitization effects are suppressed. 
Still, scanning for scientific use should be 
conducted with 16 bit sampling, which is the norm for raw CCD data produced by 
many observatories.
 
	One side of each plate envelope contains pertinent 
information about the observations, and these were 
also scanned. Ideally, the envelope images would be coupled with 
each scanned spectrum in an archive. While not discussed here, a digital 
copy of observing logs that are not restricted to the date of 
observation would also be extremely useful to assist in the 
identification of comparison stars and information that 
may not be recorded on each page of the log, such as plate and emulsion type.

	We have examined the uniformity of the illumination 
across the $8.5 \times 11$ inch surface that abuts the upper and left hand 
edges of the scanning surface. A single plate was positioned at various 
points on this surface and then scanned. Light levels were then 
measured in the same parts of each extracted plate image. A comparison 
of the light levels reveals a dispersion in the measurements that is $< 1\%$ 
of the signal level. The largest deviations are 
associated with the left hand edge of the scanning 
surface, and the dispersion for plates that do not contact this edge 
is substantially lower, amounting to $\sim 0.2\%$. These comparisons suggest 
that plates should be offset from the edges of the scanning surface.

	An example of scanned spectra and their protective envelopes is 
shown in Figure 1. The spectra and calibration 
information occupy only a modest fraction of each plate. However, as discussed 
in the next section, information from a large part of the scanned plate 
is used during the initial stages of processing. Non-sky artifacts 
such as minor cosmetic defects and writing are also visible on the 
plates. Unfortunately, the information on the 1925 plate 
envelope is in pencil, and has faded with time. This 
highlights the need to digitize information from that era in a timely 
manner before it degrades further.

\begin{figure}
\figurenum{1}
\epsscale{1.0}
\plotone{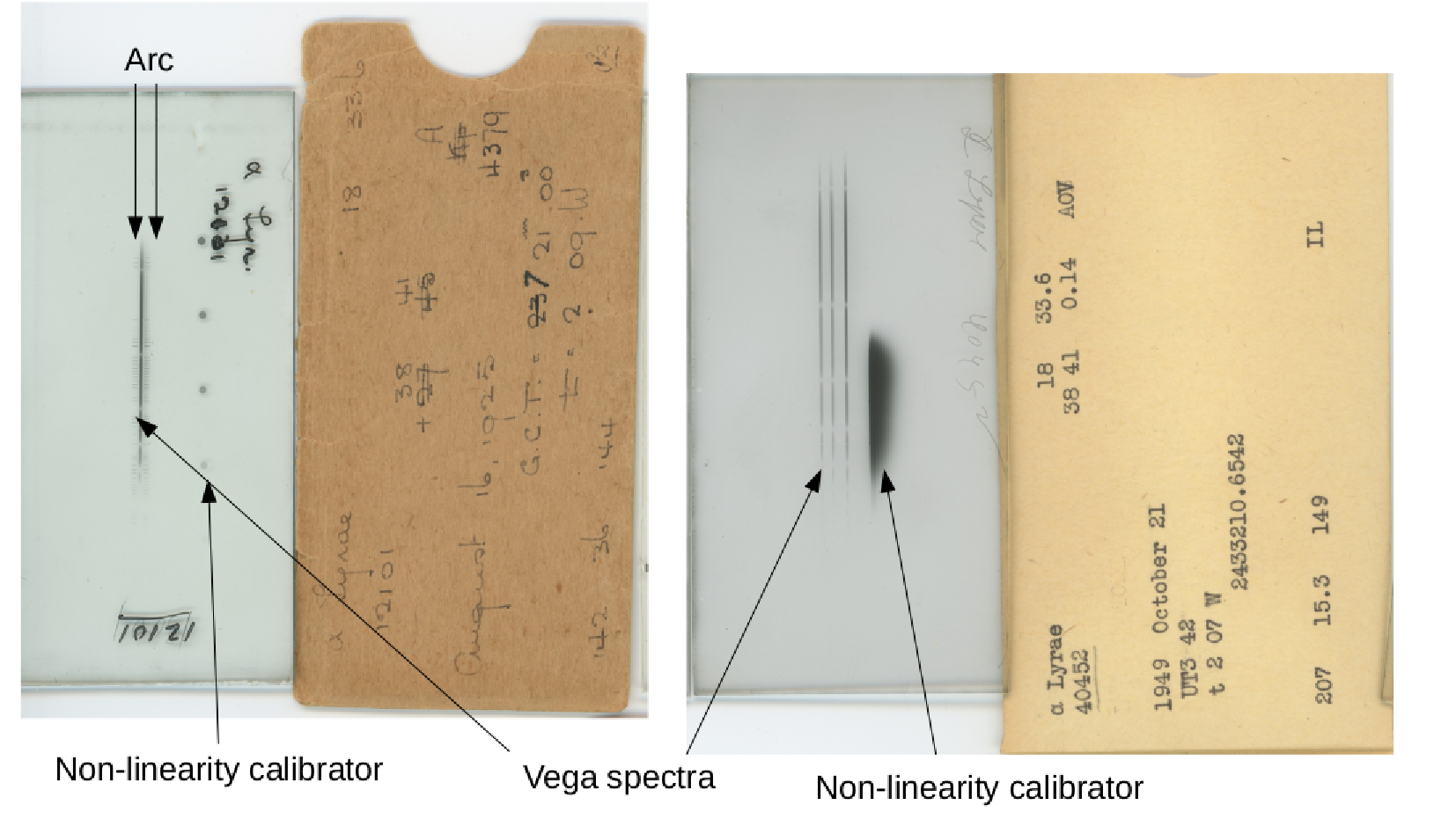}
\caption{Scanned spectra of Vega recorded on August 14, 1925 
and October 28, 1949 with the 1.8 meter telescope and the Cassegrain 
spectrograph. The front of the protective envelope for each plate is also 
shown. The plates are $10 \times 5$ cm in size. These plates show the 
range of calibration information that might be expected in the 1.8 meter 
plate collection. The stellar spectrum, emission arcs, and 
linearity calibration information are indicated. 
Arcs were not recorded with the 1949 spectra. 
The linearity calibration information is restricted to a series of 
dots to the right of the 1925 Vega spectrum. In contrast, there is 
a staircase-like linearity assessment region that parallels the 
dispersion axis of the 1949 spectra. This calibration area is saturated 
in some locations and is subject to smearing from scattered light.} 
\end{figure} 

	The spectra in Figure 1 show the range of calibration information 
in the 1.8 meter plate collection. The linearity calibration information 
for the 1925 spectrum consists of a series of dots 
with graduated intensity to the right of the 
stellar spectrum. In contrast, the 1949 plate contains three spectra of Vega 
and has a continuous staircase-like calibration region to assess response. 
As this can be traced along two dimensions then it should allow 
for the construction of a much better sampled characteristic curve than 
what could be extracted from the calibration dots in the 1925 
spectrum, although saturation and scattered light in this part of 
the 1949 plate are complicating factors. The 1949 plate lacks the 
emission arc spectra that are required for an independent 
wavelength calibration.

	All plates were scanned at 2400 dpi (i.e. 
roughly $10\mu$m spatial resolution), based on the 
results of experiments that are discussed in Section 5. 
Scanning an $8.5 \times 11$ inch area at 2400 dpi resolution with the 12000XL 
scanner in greyscale mode took just under 10 minutes from the start of 
scanning to the writing of data to disk. The scanning time 
scales roughly with the square of the scanning density. 
Therefore, scanning at 1200 dpi will proceed $\sim 4\times$ 
faster than scanning at 2400 dpi.

\section{PRODUCING A SPECTRUM}

	The scanned plates require processing to extract spectra and 
remove signatures introduced by the instrumentation, atmospheric conditions, 
and the scanning process. Basic processing steps are described below. 
These were done using tasks in the IRAF {\it images} and {\it onedspec} 
packages. These allow for an assessment of the scanned spectra while 
also demonstrating the information content of bright star spectra in the 
1.8 meter plate collection. The processing does not include corrections for 
departures from linearity, and further processing will undoubtedly be 
required for more specific applications.

\subsection{Initial Processing and Extraction}

	Scanned spectra were saved to disk in TIFF 
format. These were then converted into FITS files using 
the GNU Image Manipulation Program \footnote[1]{Copyright (C) 2000,2001,2002 
Free Software Foundation, Inc. 59 Temple Place, Suite 330, Boston, 
MA 02111-1307 USA} to permit subsequent processing with IRAF 
\citep[]{tod1986, tod1993}. The pixel format of the FITS files 
was either 8 bit or 16 bit, depending on the format 
selected for scanning (Section 3).

	The next step was to extract an area from each plate that includes 
the stellar spectrum, arc lines, calibration regions for the 
characteristic curve, and the surrounding `empty' areas. These latter 
regions are used to assess the signal produced by the scanning light source, 
and track low spatial-frequency variations in the background light level. 
Areas that include writing and other nonsky artifacts were avoided 
when possible.

	There is a tendency for the scanned images to not align 
with the cardinal axes. This could result from 
the intrinsic shape of the plate, the placement 
of the plate in the cassette at the telescope, and/or 
the positioning of the plate on the scanning surface. The 
scanned image was adjusted for this by tracing the 
ridgeline of the spectrum and applying a rotation correction. 
The 1.8 meter spectra do not show signs of curvature, 
and so a single rotation correction was adequate.
The digitized spectra were then inverted (i.e. multiplied by --1) to 
produce a positive image. 

	The blank area on each plate contains signal from the scanning light 
source. This background light level was measured in the area of each plate 
that is well separated from the stellar spectrum and the calibration 
information by fitting a linear relation to the signal on a row-by-row basis. 
These fits were made perpendicular to the dispersion axis, and the results 
were subtracted out. The fitting of a linear relation allowed 
low spatial frequency variations in the light level that 
might occur due to fogging or in scattered light to be tracked. 
The removal of the background signal in this manner produced a flatter 
background than that in the raw scanned plates. 

	An example of background subtraction is shown in Figure 2, 
where a cut through the scanned 1963 Vega plate perpendicular 
to the dispersion axis before and after background subtraction is 
shown. The section shown in Figure 2 is at a wavelength 
that is close to H$\gamma$. As would be expected from Figure 1, the background 
light is markedly fainter than the light from the 
stellar spectra, the arcs, and the linearity calibration region. 
The removal of a linear trend based on the signal in 
the areas indicated with the green lines greatly suppresses systematic trends 
in the signal. Residual systematic variations remain near the 
left hand edge after background subtraction, indicating that a 
linear relation is only an approximation of large-scale trends. Scans done 
with the V800 and 12000XL scanners show similar results, suggesting that 
background trends like that shown in Figure 2 are intrinsic to the plate, 
and not an artifact of the scanning process or scanner.

\begin{figure}
\figurenum{2}
\epsscale{1.0}
\plotone{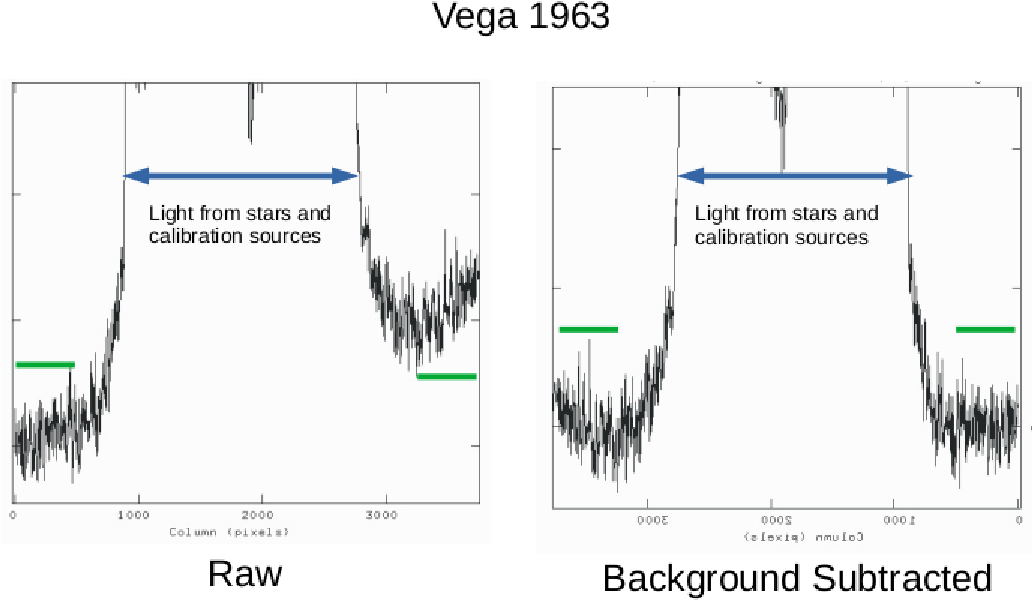}
\caption{Background light subtraction in the 1963 plate. The panels show 
a cut made perpendicular to the dispersion direction 
before and after background subtraction. This cut samples a wavelength 
near H$\gamma$. Light from the stars, arcs, and linearity 
calibration region swamp the background light in the central regions of 
the cut. The green lines mark the approximate regions used to 
fit the large-scale linear trend in background 
light that was subtracted from the data. While much 
of the variation in the background regions is suppressed with the 
removal of a linear trend, the upward-pointing signal 
in the left hand edge of the background-subtracted 
panel indicates that a linear fit to the background is only an approximation.}
\end{figure} 

	Lines in the emission arc were then extracted. This was done 
by co-adding the signal across each of the arc spectra that bracket 
the stellar spectrum. These arcs were presumably 
recorded before and after the science exposure to account for instrument 
flexure. The arcs do not monitor the wavelength 
resolution of the stellar spectra as the arc light is not fed through 
the main slit. There is a $\sim 10 - 20\mu$m (i.e. 
1 -- 2 resolution elements with 2400 dpi scanning) offset 
between line locations in the two arcs perpendicular to the dispersion axis. 
Such small offsets are consistent with an expectation of minimal flexure 
given the short exposure times associated with observing a bright source 
like Vega, and also validate the rotation correction applied to 
the scanned images.

	The extraction of a steller spectrum is more 
complicated than simply co-adding the signal along the slit. 
This is because the spectral resolution in these spectra varies in a 
systematic way with location along the slit. We suspect that this is 
because the star was not trailed along the slit given the short exposure time 
required for such a bright source. The difference in 
resolution is substantial, and this is demonstrated in 
Figure 3, where a section of the 1925 spectrum centered on 
H$\gamma$ is shown. Spectra extracted from three locations along the slit are  
displayed, and there are obvious differences in the appearance of H$\gamma$. 
This behaviour is seen in all of the spectra examined here.

\begin{figure}
\figurenum{3}
\epsscale{1.0}
\plotone{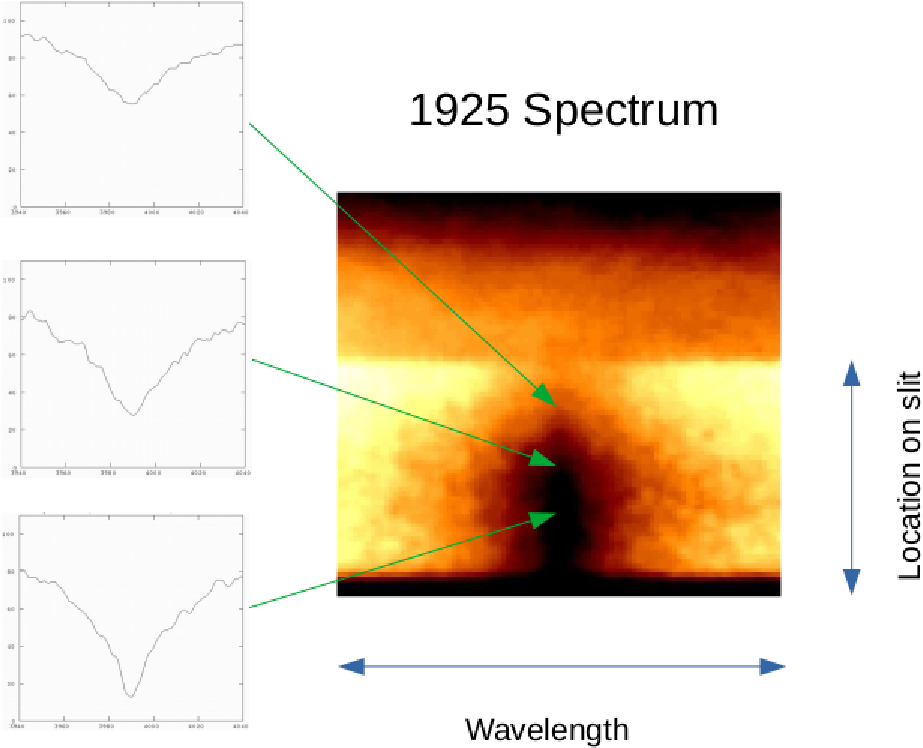}
\caption{Wavelength resolution along the slit in 1.8 meter telescope
Cassegrain spectra of Vega. The right hand image shows a section of the 
scanned 1925 Vega spectrum centered on H$\gamma$; the light 
is dispersed along the vertical axis, while the horizontal axis runs along 
the slit. Spectra at three locations along the slit are shown to the left, and 
the change in spectral resolution with location along the slit can be seen. 
This demonstrates the need to capture the full spectrum in two dimensions when 
scanning a 1.8 meter plate in order to identify an optimum extraction region.}
\end{figure}

	There are a number of options for extracting stellar spectra from 
these data, and the criteria adopted for any given 
study will depend on requirements placed on 
wavelength resolution, the wavelength interval to be sampled, S/N 
ratio, etc as defined by the science goals. The flexibility 
required when extracting spectra reinforces the 
need for the slit profile to be scanned in its entirety for archival purposes. 
For the current study, we have adopted what we consider to be generic 
extraction criteria. The extraction region was identified as a compromise 
between S/N, line sharpness, and linearity.

	Spectra were extracted such that the continuum and weak lines 
near the peak in the wavelength response appear to fall along 
a linear part of the characteristic curve. To do 
this, spectra at various offsets from the central ridgeline were 
visually inspected and a spatial interval along the slit was identified 
where the signal is high and varies in a more-or-less linear manner 
from column to column along the slit profile to within a few percent 
near the continuum. This assessment was done at wavelengths bracketing 
H$\gamma$. Multiple spectra of Vega were recorded on the 1949 and 1963 
plates, and individual spectra were extracted from these plates. 
This allows the comparison of spectra taken from the same plate but with 
different observing times (e.g. as in the case of the 1949 spectra -- Table 1). 

\subsection{Comparing Partially Processed Spectra}

	The spectra discussed in this paper were recorded over a four 
decade time span, and during this time there were changes to the 
spectrograph, the properties of photographic 
materials, hypersensitization methods, and observing 
techniques. The spectra examined here thus are expected to 
show a diverse range of properties. The extracted spectra are 
compared in Figure 4, where the signal level 
is in uncalibrated digital units (hereafter 'DU').
The 1949 and 1963 spectra are the means of all spectra on those plates.

\begin{figure}
\figurenum{4}
\epsscale{1.0}
\plotone{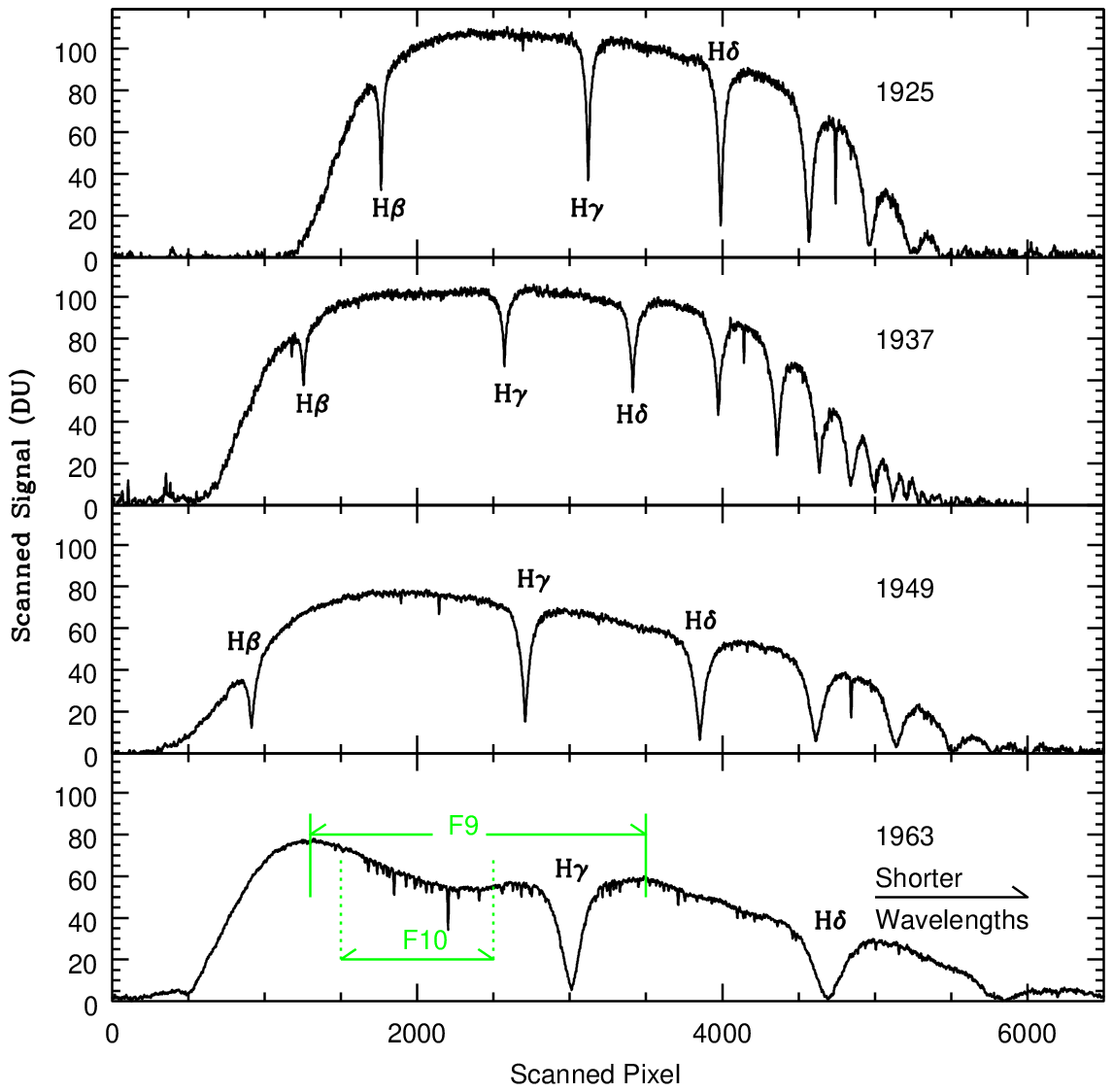}
\caption{Partially processed spectra digitized with the Epson 12000XL 
scanner at 2400 dpi. The vertical scale shows uncalibrated digital units, 
while the horizontal axis shows pixels in the 
scanned image, with wavelength decreasing towards the right. 
Pixels span $9.5\mu$m on a plate at this scanning density. 
The 1949 and 1963 spectra are the means of 
the spectra that were recorded on each plate. Epoch-to-epoch 
differences in wavelength coverage, wavelength response, wavelength 
resolution, and S/N are clearly evident. The 1963 spectrum has the 
highest wavelength resolution and S/N, and numerous metallic absorption 
features are clearly seen in that spectrum. The wavelength intervals in the 
1963 spectrum that are compared with a CCD spectrum in Figures 11 and 12 
are indicated.} 
\end{figure}

	There are obvious differences in overall throughput 
and spectral resolution among the spectra. H$\gamma$ and 
H$\delta$ tend to be the deepest features. While other Balmer 
lines are also present, in many cases these are in areas of low throughput. 
The wavelength coverage of the 1937 spectrum extends to the shortest 
wavelengths, although the throughput at wavelengths that sample the higher 
order Balmer lines is low. Another interesting aspect of 
the 1937 spectrum is that the Balmer lines are much weaker 
than in the other spectra. The cause of this discrepancy in line 
strength is a matter of speculation, although 
the 1937 spectrum was recorded at the end of night during twilight, and 
scattered light could affect line depths. Despite 
the discrepant depths of the Balmer lines, we include the 1937 spectrum to 
demonstrate the range of spectra that are in the DAO collection.

	The 1925 and 1937 spectra have roughly the same wavelength dispersion, 
whereas the 1963 spectrum has the highest spectral resolution 
and S/N. This is likely why there is a rich population of moderately 
weak absorption features in the 1963 spectrum that are not 
clearly seen in the 1925 and 1937 spectra. Such weak lines are of interest 
as their shallow nature increases the chances that their 
entire profiles will sample linear portions of the characteristic curve, 
which may not be the case for deep lines.

\clearpage
\subsection{Wavelength Calibration and Normalization}

	The arcs that bracket each spectrum were 
aligned based on the centroids of emission lines, and 
then averaged together to produce a final arc. 
The shifts applied to the arcs were of a size that 
did not significantly increase the intrinsic widths of the emission lines 
in the final arc. Arcs were not recorded for the 1949 spectra, and the 
wavelength calibration for those spectra is based on the centers 
of the Balmer lines. 

	Prominent Fe emission lines were identified in each mean arc spectrum 
and a polynomial fit was made to relate line location 
with wavelength. The resulting dispersion 
solution was then applied to the stellar spectrum. 
The final processing step was to normalize each spectrum to a 
pseudo-continuum. This was done by fitting a polynomial 
to parts of each spectrum that are free of deep absorption features, 
and then dividing by the result.

\section{SCANNING EXPERIMENTS}

\subsection{Scan Density}

	The sharpness of lines in a digitized spectrum will progressively 
improve with increasing scan density until the intrinsic resolution limits 
imposed by the spectrograph optics and/or the MTF of the scanner are/is 
reached. While one approach might be to scan at the highest possible 
sampling density, this may prove to be inefficient 
as the time to digitize a plate depends on the 
scanning density; this might be a critical consideration if digitizing a 
large collection of plates. Moreover, the delivered scan density may be 
compromised by the MTF of the device -- scanning at the highest possible 
density might produce no gain in quality at the expense of much longer 
digitization times. Tests were thus conducted to determine a preferred 
scanning density. 

	An initial series of tests were performed on the 
1925 Vega spectrum with the Epson V800 scanner. 
Portions of the 1925 spectra centered on H$\gamma$ 
were digitized at densities of 300, 1200, and 2400 dpi, and the results are 
compared in Figure 5. The spectra in this figure were extracted over the 
same spatial intervals along the slit. Each spectrum was wavelength calibrated 
using the arc obtained from the 2400 dpi scan. This required interpolation 
within the 300 dpi and 1200 dpi spectra to match the sampling of 
the 2400 dpi spectrum.

\begin{figure}
\figurenum{5}
\epsscale{1.0}
\plotone{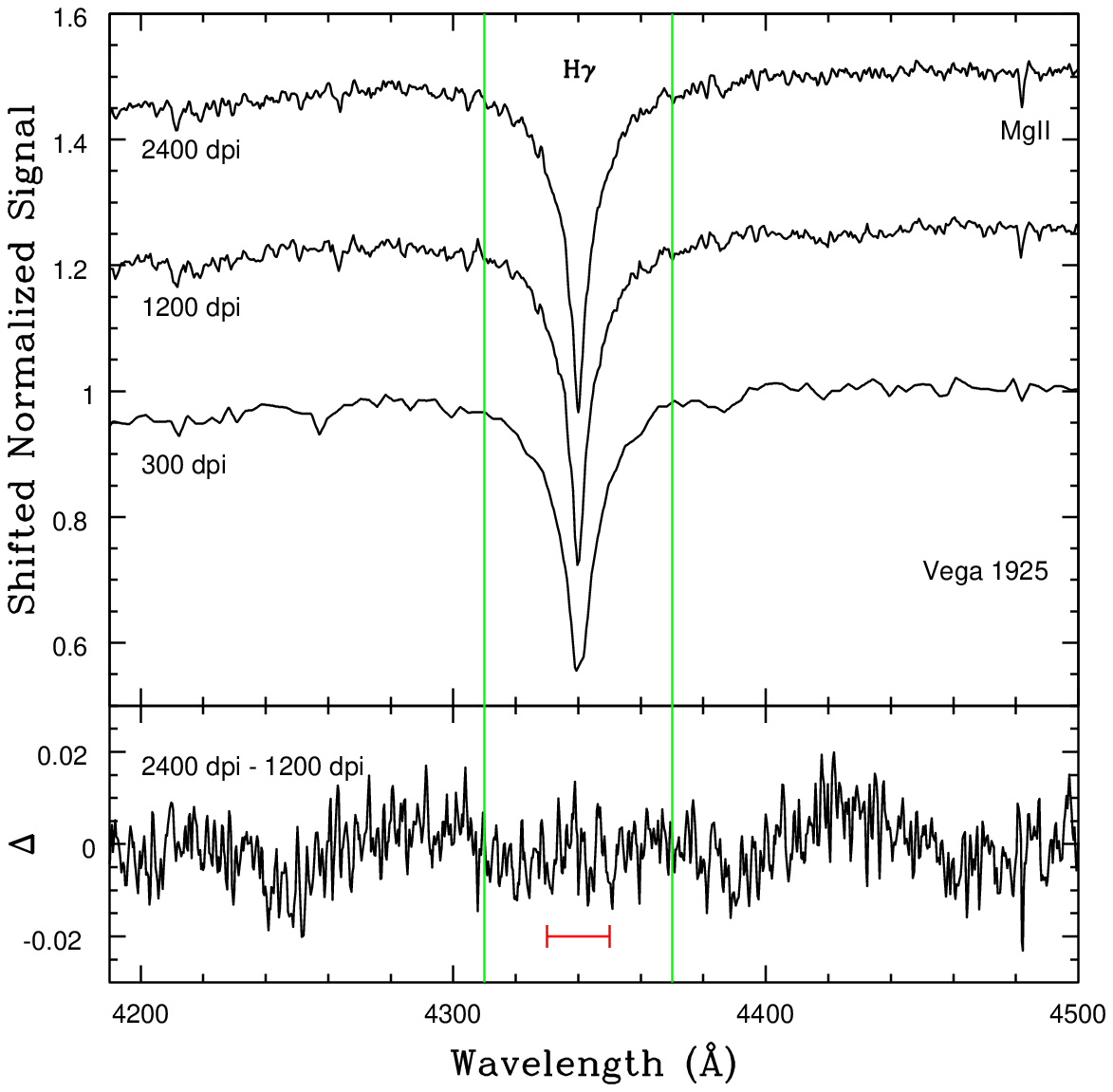}
\caption{(Top panel:) Spectra of Vega recorded in 1925 and digitized 
at densities of 300, 1200, and 2400 dpi with the Epson V800 scanner. 
The vertical green lines mark the approximate limits of H$\gamma$, 
and the loaction of MgII 4481 is also marked. 
There is a clear difference in resolution 
between the spectra digitized at 300 and 1200 dpi. 
However, the differences between the spectra scanned at 1200 and 2400 
dpi are more subtle, with weak features, such as 
MgII 4481, appearing to be slightly better 
defined in the 2400 dpi spectrum. (Bottom panel:) The differences between 
the 2400 dpi and 1200 dpi spectra, where $\Delta$ is the difference 
in the sense 2400 dpi -- 1200 dpi. There is a high frequency component 
with a $\sim \pm 0.05\%$ dispersion that is superimposed on a 
lower frequency modulation with a $\pm 1\%$ amplitude. The former 
is likely due to noise, while the latter has a peak-to-peak distance 
that is roughly twice the width of H$\gamma$. 
There is a $\sim 0.5\%$ difference between the two spectra near H$\gamma$, 
in the sense that the central regions of H$\gamma$ may be slightly 
sharper (but not deeper) in the 2400 dpi spectrum. The red line marks the 
approximate wavelength limits of this peak. As for MgII 4481, 
this feature is clearly deeper in the 2400 dpi spectrum. This comparison 
suggests the use of a scan density of at least 1200 dpi when using the 
V800 scanner, with a possible preference for 2400 dpi.}
\end{figure} 

	The depth and width of H$\gamma$ 
changes noticeably between scan densities of 300 and 1200 
dpi. However, the differences between the 1200 and 2400 dpi spectra are 
more subtle, and the difference between the 1200 
dpi and 2400 dpi spectra is shown in the bottom panel of Figure 5. 
There are high frequency variations with a dispersion of $\pm 0.05\%$, 
and the random nature of this component is suggestive of  
noise. There is also a low frequency component 
that has an amplitude of $\pm 1\%$ and a peak-to-peak wavelength 
of $\sim 70\AA$. We are uncertain as to the origin of this modulation, 
but note that it has a periodicity that is $\sim 2\times$ the width of 
H$\gamma$.

	There is a poorly defined peak in the lower panel 
that is centered on H$\gamma$. This feature 
has an approximate width of $\sim 30\AA$ and an amplitude of $\sim 0.05\%$; 
the location of this feature is indicated with 
the red line. This peak is in the sense that the core of H$\gamma$ may be 
slightly sharper, but not deeper, in the 2400 dpi spectrum. The residuals in 
the lower panel indicate that MgII 4481 is deeper in the spectrum 
scanned at 2400 dpi.

	The comparisons in Figure 5 indicate that while 
a scan density of 1200 dpi with the V800 
appears to detect most features, there is possible improvement when 
scanning at 2400 dpi. These comparisons thus argue for a scanning density 
{\it with the V800 scanner} of at least 1200 dpi, with a preference for 2400 
dpi. We note that sampling at 2400 dpi produces pixels with sizes near the 
lower end of the 10 -- $30\mu$m resolution of photographic emulsions, and 
that are finer than the $15\mu$m pixels used in many CCDs. 

	Photographic emulsions evolved over the last century, and the 
grain sizes in the 1925 and 1937 plates are likely larger than those used 
in more recent plates. To assess if this affects the choice of scanning 
density, the 1963 plate was scanned at both 2400 dpi and 3600 dpi 
with the 12000XL scanner, and the results for one of the spectra 
on the 1963 plate are shown in Figure 6. As with the comparisons in 
Figure 5, the spectra in Figure 6 were extracted from the same spatial 
interval along the slit.

\begin{figure}
\figurenum{6}
\epsscale{1.0}
\plotone{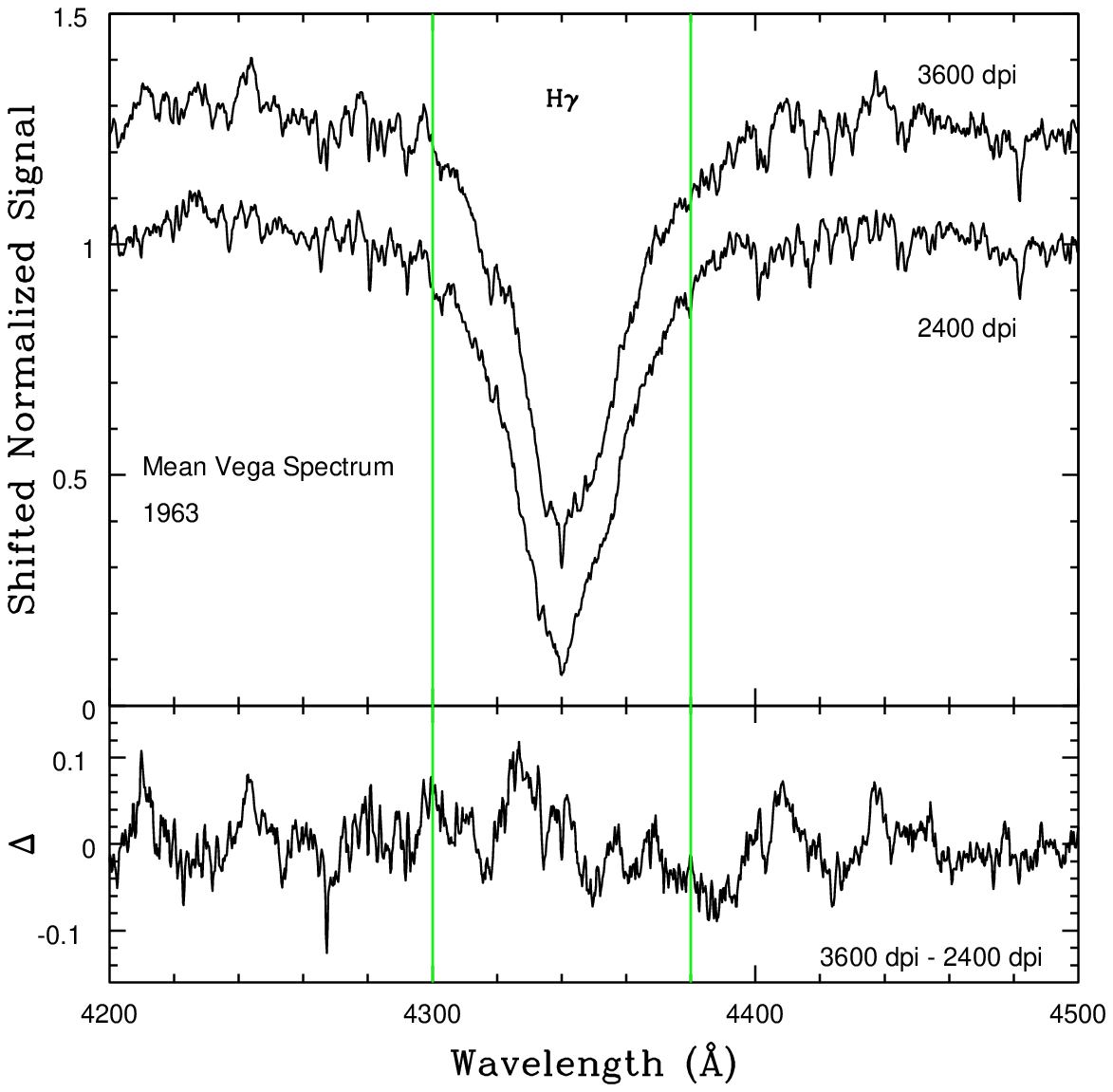}
\caption{(Top panel:) Spectra of Vega recorded in 1963 and digitized at 2400 
and 3600 dpi with the Epson 12000XL scanner. There is substantially more 
jitter in the 3600 dpi H$\gamma$ profile than in the 2400 dpi profile, and 
this is likely due to scanning at too high a density. This jitter is clearly 
seen near the core of H$\gamma$ in the 3600 dpi spectrum. (Bottom panel:) 
The difference between the two spectra. The difference near H$\gamma$ is 
tilted, suggesting that the 2400 and 3600 dpi spectra have slightly different 
behaviour in the short wavelength shoulder of this line. Based on the 
comparisons shown in this figure and Figure 5, 
2400 dpi is adopted as the default scanning density.} 
\end{figure} 

	There are clear differences between the spectra in Figure 6. 
The spectrum scanned at 3600 dpi has noticeably more 
jitter than the 2400 dpi spectrum. Moreover, the core of H$\gamma$ 
is also flatter at 3600 dpi, although there is a narrow downward-pointing 
spike at the line center. That the difference between the spectra in 
the lower panel is tilted at wavelengths near H$\gamma$ suggests that 
the spectra have different profiles, and this is most evident in the line 
shoulder at shorter wavelengths. In Section 6.4 it is demonstrated that the 
symmetric line profile of the spectrum scanned at 2400 dpi is consistent 
with the shape of the line profile in CCD spectra.

	We suspect that the ragged appearance of the 3600 dpi spectrum 
in Figure 6 is a consequence of sampling at too high of a density, such that 
noise is introduced into the scanned spectrum due to the lower signal 
per resolution element. The jagged appearance of the central 
regions of H$\gamma$ when scanned at 3600 dpi is an obvious noise 
signature, and CCD spectra with similar wavelength resolutions 
show a smoothly varying profile near the line core (Section 6.4). 
Based on the comparisons in Figures 5 and 6, 
we adopt a scanning density for the 12000XL of 2400 dpi for the 
remainder of this study, as this sampling appears to resolve 
most features without contributing additional noise.

	We close the discussion on scan density by noting that the V800 
scanner offers a `full auto' scanning mode, for which the scan density and 
gain are set automatically. The 1925 spectrum was scanned in this 
mode, and the results were poor: the brightest parts of 
the spectrum were saturated, while the delivered resolution appears 
to be $\sim 300$ dpi, which is much lower than that required to fully resolve 
information in the spectrum. Thus, we do not recommend this setting 
to digitize spectra.

\clearpage
\subsection{Comparing Scanners}

	The V800 and 12000XL scanners have different optical components 
and -- possibly -- detectors \footnote[1]{We were unable to obtain 
specific information about the detectors in either device from EPSON support 
services.}. Differences in the optics have the potential to affect the MTFs of 
the devices, such that one device might deliver better angular resolution 
than the other at higher scan densities. Differences in the MTF may also 
introduce subtle effects in light scattered from the 
illumination source in the scanner.

	The spectra recorded in 1925 were digitized with both devices 
at a density of 2400 dpi, and the results 
are compared in Figure 7. The result of subtracting the 12000XL spectrum from 
the V800 spectrum is shown in the bottom panel, and clear differences are seen. 
When compared with the V800 spectrum, H$\gamma$ is narrower and 
$\sim 20\%$ deeper in the spectrum digitized with the 12000XL scanner. 
Weak features are also sharper and better defined in the 12000XL spectrum. 
There is a $\sim \pm 1\%$ dispersion in the difference on either side of 
H$\gamma$, and there is an absence of the low frequency trends seen 
in Figure 5. Both devices thus traced the 
continuum in a consistent manner to within $\sim 1\%$. 

\begin{figure}
\figurenum{7}
\epsscale{1.0}
\plotone{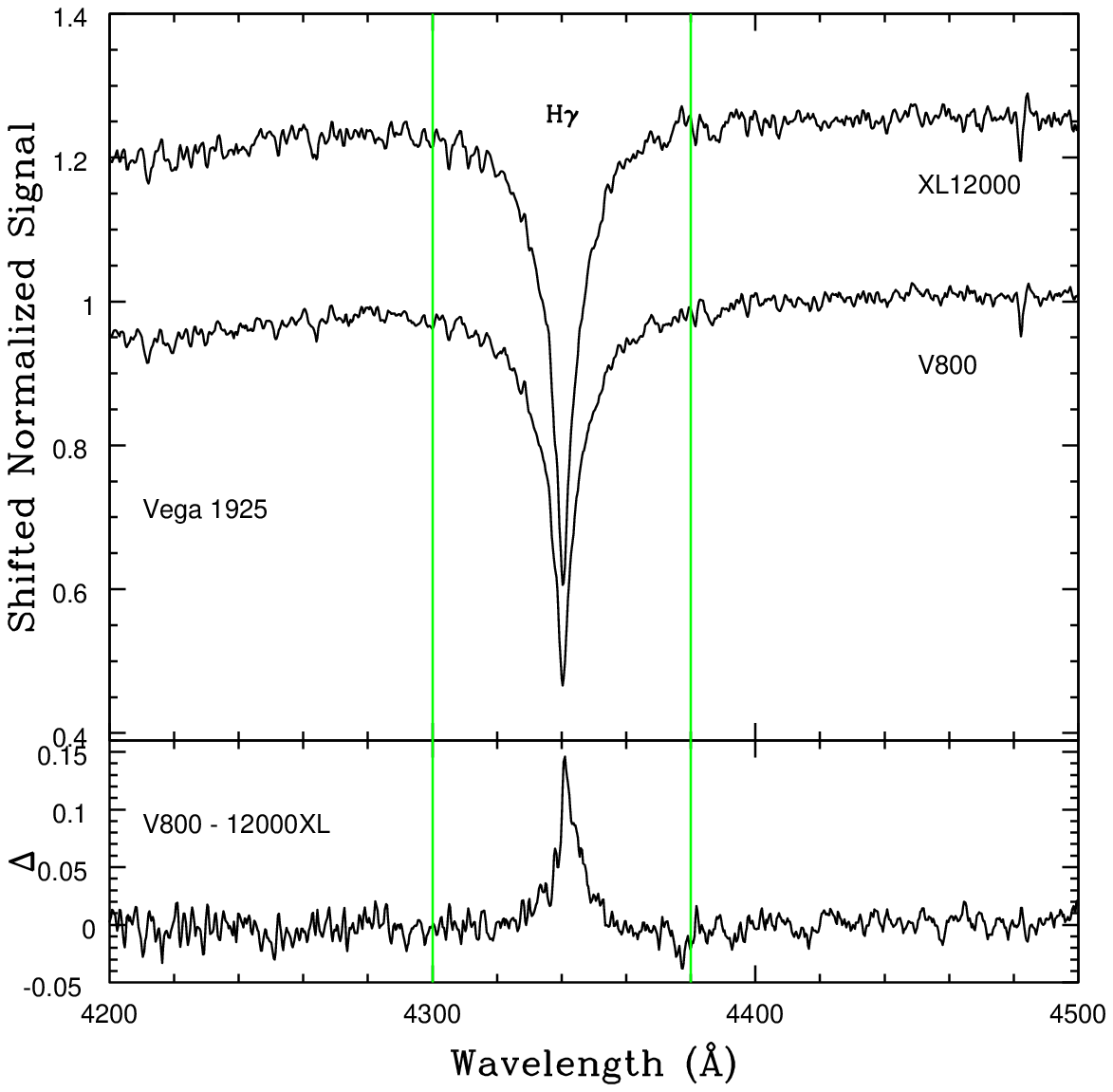}
\caption{(Top panel) Spectra of Vega recorded in 1925 
and digitized at 2400 dpi with the Epson 12000XL and V800 
scanners. (Lower panel) The difference between these spectra ($\Delta$), 
in the sense V800 -- 12000XL. H$\gamma$ is narrower and $\sim 20\%$ deeper 
in the 12000XL spectrum when compared with the V800 spectrum. 
This is due to differences in the scanner MTFs, in the sense that the 12000XL 
scanner has superior angular resolution when scanning at 2400 dpi. 
The 12000XL is thus prefered for the scanning of spectra.} 
\end{figure} 

	The differences between the spectra in Figure 7 are likely due
to the MTFs of the scanners, in the sense that the MTF of the 
12000XL is superior to that of the V800 at 2400 dpi scanning densities. If,
as suggested in Figure 7, the MTF of the V800 scanner blurs the signal 
at densities higher than 1200 dpi, then the comparison in Figure 5
is affected, as the modest differences between the 1200 and 2400 dpi spectra 
are then a consequence of the V800 MTF significantly smearing the signal 
at resolutions in excess of 1200 dpi. The differences between the 1200 
dpi and 2400 dpi spectra would then be much larger if the MTF of 
the V800 were similar to that of the 12000XL. Given the evidence for 
a difference in the MTFs then we will focus exclusively on spectra 
digitized with the 12000XL at densities of 2400 dpi for the remainder 
of the paper.

\section{COMPARING SPECTRA}

\subsection{Comparing the Processed Spectra}

	The scanned spectra in Figure 4 show 
a diverse range of characteristics. To further investigate 
any differences and similarities we examine processed versions of these 
spectra. We focus on the wavelength interval near H$\gamma$, as the spectra 
at all four epochs are more-or-less centered near this feature. 

	The processed spectra are compared in Figure 8. 
H$\gamma$ in the 1963 spectrum differs from that in the other epochs, in 
that it is broader and deeper, and so has a larger equivalent width. 
In contrast, H$\gamma$ and the other Balmer lines in the 1937 
spectrum are systematically shallower than in the spectra at other epochs.
H$\gamma$ is the feature in Figure 8 that may be most susceptible 
to departures from linearity, given that it is 
deep, and so samples the widest range in signal.

\begin{figure}
\figurenum{8}
\epsscale{1.0}
\plotone{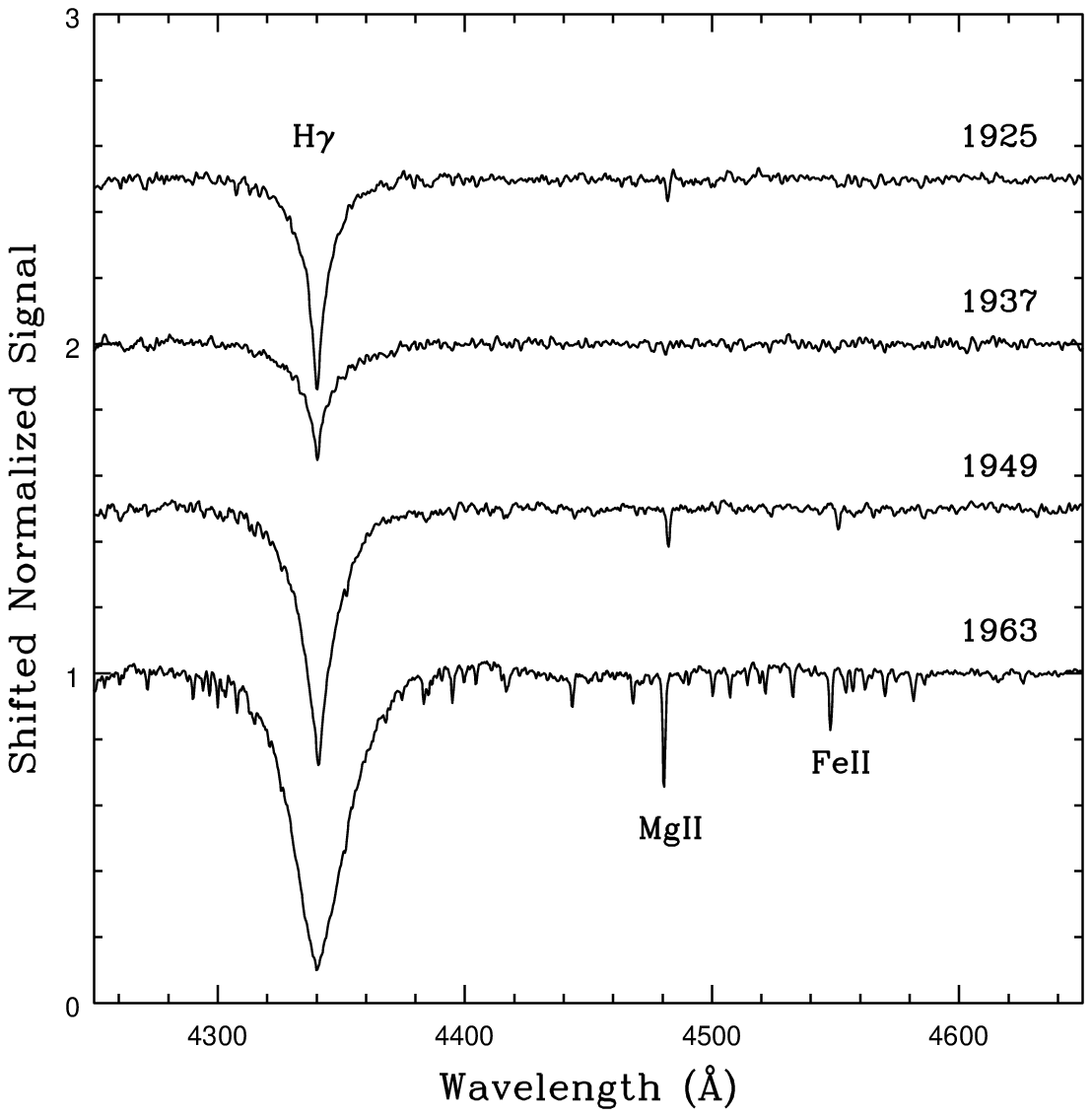}
\caption{Processed spectra. The 1949 and 1963 
spectra are the means of the multiple spectra recorded 
on each plate. The spectra have different wavelength resolutions and 
response characteristics (e.g. Figure 4). Despite having similar 
wavelength coverage and wavelength dispersion, the depth of H$\gamma$ 
in the 1925 and 1937 spectra clearly differs, and this may be due to the 
latter having been recorded during twilight. Numerous absorption 
features are apparent between 4400 and 4600\AA\ in the 1949 and 1963 
spectra, and the detection of these is likely due 
to the higher wavelength resolution of those spectra.} 
\end{figure}

	H$\gamma$ aside, the characteristics of the spectra 
from 1925 to 1963 differ in the sense that the 
absorption features between 4400 and 4600\AA\ that are 
prominent in the 1963 spectrum are not seen in the 1925 and 1937 spectra. The 
1949 spectrum is an intermediate case, in that weak features are apparent in 
that spectrum, but with a poorer S/N than in the 1963 spectrum. 
These features are also weaker in the 1949 spectrum than in the 1963 spectrum. 

	The MgII 4481 line is clearly detected 
in three of the spectra, and there is a weak feature at this 
wavelength in the 1937 spectrum. The relative depth of MgII with respect to 
H$\gamma$ varies from spectrum-to-spectrum, and this may be due to differences 
in the characteristic curves of the plates. Early experiments with 
hypersensitization at the DAO were not entirely 
successful \citep[e.g.][]{pla1921}, leading us to suspect 
that the 1925 plate was not hypersensitized. There is also no mention of 
hypersensitization in the log entry for the 1937 plate. However, the 1949 
plate (and presumably the 1963 plate) was hypersensitized. We also 
note that the 1925 and 1937 spectra have a lower dispersion than the 1963 
spectrum (e.g. Figure 4), whereas the dispersion of the 1949 spectrum is 
lower than that of the 1963 spectrum, but higher than that of the 1925 and 
1937 spectra. Therefore, the observed range in properties of the features 
between 4400 and 4600\AA\ is also affected by wavelength resolution. 

\subsection{Intraplate Comparisons}

	Differences in spectra that are recorded over long time scales, 
like those discussed in Section 6.1, are to be expected. However, 
differences in line profiles can also occur among 1.8 meter slit spectra 
recorded on the same night. Comparisons of spectra recorded on the 
same plate are of particular interest as these have 
the same emulsion batch and spectrograph configuration. To the extent 
that scanning captures the features in a spectrum 
then differences between spectra recorded on the same plate might then 
be attributed to nonlinearity and/or variations in the seeing, both of
which can affect the shape of line profiles in slit spectra. 
Comparisons of such spectra then illustrate the intrinsic level of 
uncertainty that might be expected in spectra like those recorded in the 
early days of the 1.8 meter telescope, and provide limits 
on what might be expected when making comparisons with spectra recorded 
with more modern techniques, such as via a fiber feed or an image slicer.

	Two spectra from the 1949 plate were selected to examine 
the stability of line characteristics. The 1963 plate also hosts multiple 
spectra. However, the 1949 plate was selected as arcs were not recorded 
on that plate. While an obvious problem for wavelength 
calibration, the absence of arcs removes the 
potential for scattered light from the arc to skew 
any comparison -- differences between the spectra on the 1949 plate are 
then likely intrinsic to the observing environment. 

	The spectra on the 1949 plate have signal levels 
that differ by roughly a factor of two. A comparison 
with Figure 4 shows that the three Balmer lines fall in parts of the 1949 
spectrum that sample different system throughputs. H$\beta$ might be the 
most susceptible of these to nonlinearity effects, as it is an intrinsically 
deep feature that falls where the overall throughput is rapidly declining with 
wavelength. A 0.3 dex difference in signal level is not that large in the 
context of the characteristic curve unless it is near the saturation 
limit (Section 6.3), leading us to suspect that nonlinearity 
may not play a {\it dominant} role in differences in line properties. 

	Two of the normalized spectra extracted from the 1949 plate 
are compared in the top panel of Figure 9. The seeing 
profile of the 1949c spectrum is $\sim 10\%$ more compact than 
that of the 1949b spectrum, and seeing might be expected to influence 
line properties for a spectrograph with a bare slit entrance aperture.
There are differences in the continuum at the $\sim 3\%$ level, while 
the Balmer lines in the 1949b spectrum in the top panel 
are deeper than those in the 1949c spectrum. The widths of the 
Balmer lines in the two spectra do not appear to be greatly different. 
The differences between the spectra are much more subdued than those 
evident among the spectra in Figure 4.

\begin{figure}
\figurenum{9}
\epsscale{1.0}
\plotone{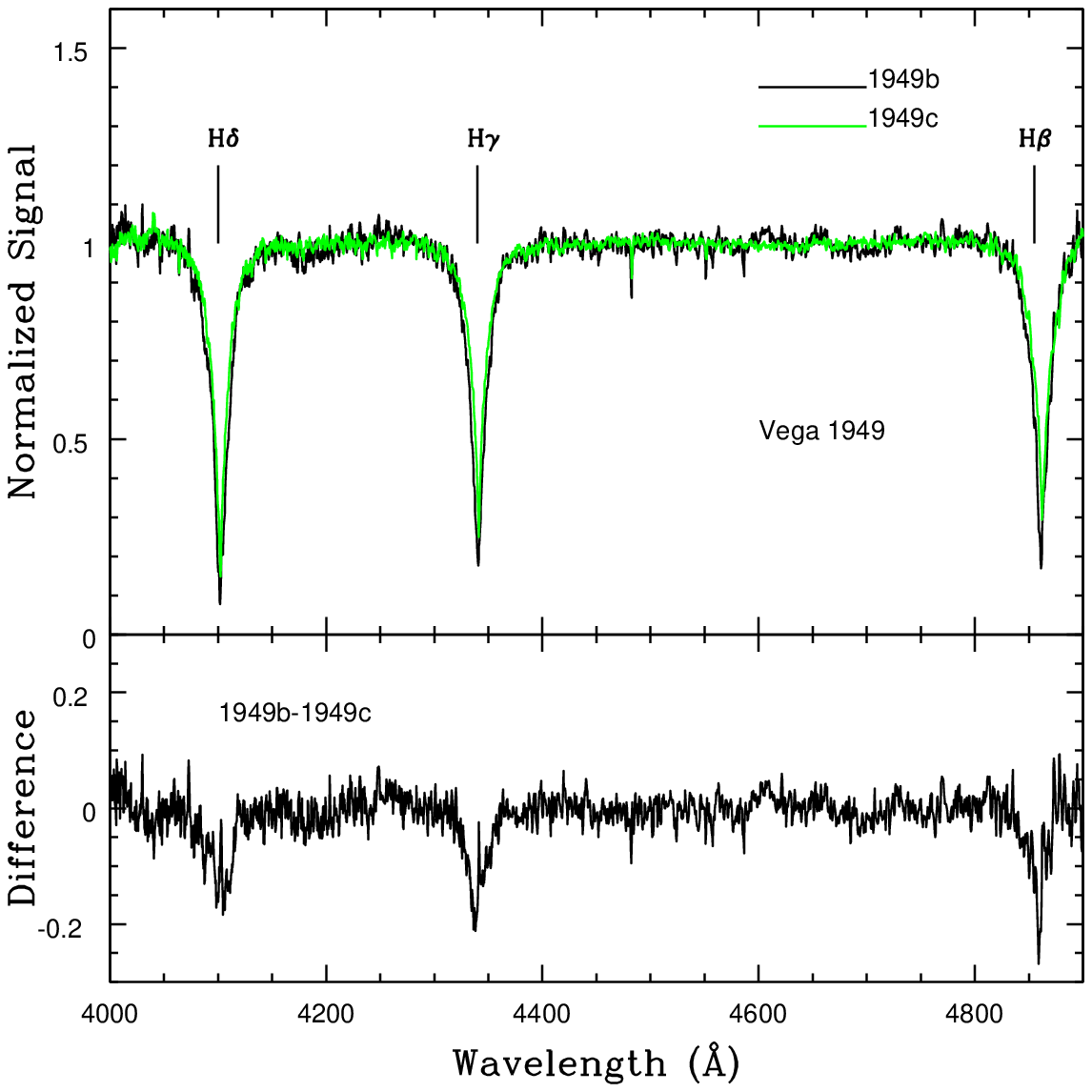}
\caption{(Top panel:) Normalized spectra of Vega from the 1949 
plate. The Balmer lines in the two spectra differ in two ways. First, 
the lines in 1949b spectrum are slightly deeper than in the 
1949b spectrum. Second, the Balmer lines in the 1949c 
spectra are slightly narrower than in the 1949b spectrum, although this 
likely reflects the difference in line depths. (Bottom panel:) The 
difference between the normalized spectra, in the sense 1949b -- 1949c. The 
depths of the Balmer lines in the two spectra differ at the $15 - 20\%$ level. 
The differences near H$\beta$ and H$\gamma$ are comparable, even though the 
former is observed at a wavelength with lower overall throughput than the 
latter. This comparison demonstrates that one might expect differences in line 
properties among spectra in the 1.8 meter collection that were recorded on the 
same plate and close together in time. These differences also demonstrate 
the magnitude of uncertainties that might be expected when making comparisons 
with spectra recorded with more modern techniques, as is done in Section 6.4.}
\end{figure}

	The difference between the normalized spectra is 
shown in the bottom panel of Figure 9. Differences in line depth between 
the two spectra are clearly evident, with the depths of the 
Balmer lines in the two spectra differing at the $15 - 20\%$ level. 
The residuals for H$\beta$ and H$\gamma$ are comparable, and 
larger than the residuals near H$\delta$. 

	We suspect that there is likely no single explanation for 
the differences seen in Figure 9. That the Balmer lines in the 
1949b spectrum are consistently deeper than those in the 1949c 
spectrum is perhaps unexpected if nonlinearity is the source of 
any differences. Subtle changes in seeing might play a role, 
although the similarity in line shape that is evident in the upper panel 
of Figure 9 argues that seeing may not be the dominant factor. 
The range over which spectra are extracted along the slit may also 
introduce differences, given the manner with which the line 
profile can vary along the slit (e.g. Figure 3). That there are inherent 
limitations to the use of bare slit spectra in the study of line profiles 
is hardly a novel finding. Still, the differences 
in line depth in Figure 9 indicate that significant uncertainties remain 
among spectra extracted from plates in the DAO 1.8 meter collection 
even if they were recorded with similar instrumentation and during the 
same observing period.

\subsection{Assessing Linearity using On-plate Information}

	A characteristic curve can be divided into three broad components: 
a shallow relation at low signals, a steeper linear relation at 
intermediate signal levels, and another shallow relation near saturation. 
Applying a comprehensive and robust correction for nonlinearity is 
beyond the scope of this study. In this section we set a more modest 
goal of identifying the range in signal in which the response appears to be 
linear. This opens the possibility of identifying parts of a spectrum 
in which line characteristics can be extracted after applying the 
basic processing steps described in Section 4.

	While nonlinearity may not be a problem 
for measuring line positions, it affects efforts to assess line 
shapes and measure equivalent widths, especially in deep lines where 
the signal may span different parts of the characteristic curve. 
On the other hand, if there is a linear response over parts 
of the recorded signal at or near the level of the 
continuum, then shallow (weak) lines will be less susceptible 
to departures from linearity, as the difference in the light levels 
between the continuum and the line cores may be modest. If this 
is the case then shallow lines might be features that 
can be studied most robustly in photographic spectra.

	Deviations from linearity can be assessed using calibration 
information that is recorded on most of the DAO plates. The mechanics of 
how the linearity calibration signal was originally fed to the slit environment 
on the 1.8 meter telescope was discussed by \citet{pla1923}. As this 
calibration information is a key part of the scientific content on the plates, 
we examine here if the information in the calibration region can be recovered 
in a useable form from plates digitized with the 12000XL scanner. The 
1963 plate was selected for this assessment as the extracted spectrum has the 
highest S/N and finest wavelength resolution of the datasets 
examined here. The 1963 spectrum is compared with a CCD spectrum in Section 6.4.

	Information extracted from the characteristic curve calibration 
area on the 1963 spectra is shown in the top panel of Figure 10, where the 
mean signal in four different columns that run perpendicular to the 
dispersion axis is shown. Each of these columns has a width of 10.5 mm 
on the plate, and light levels in five steps 
are sampled. The curves in the figure are labeled 
C1 -- C4. The signal in a large part of the calibration region is saturated, 
and the measurements in Figure 10 were extracted from parts of the 
calibration region where the signal is not saturated over the full range 
of five steps in each column, although the signal in Step 5 of C1 is 
close to saturation. The stepped nature of the calibration curves is 
blurred by scattered light and the Eberhard effect, 
in which edges between high contrast areas on the plate are accentuated if 
the developing solution is not smoothly mixed across the plate 
\citep[][]{ebe1926}. This blurring, combined with scattered light, is a 
source of uncertainty when assessing linearity.

\begin{figure}
\figurenum{10}
\epsscale{1.0}
\plotone{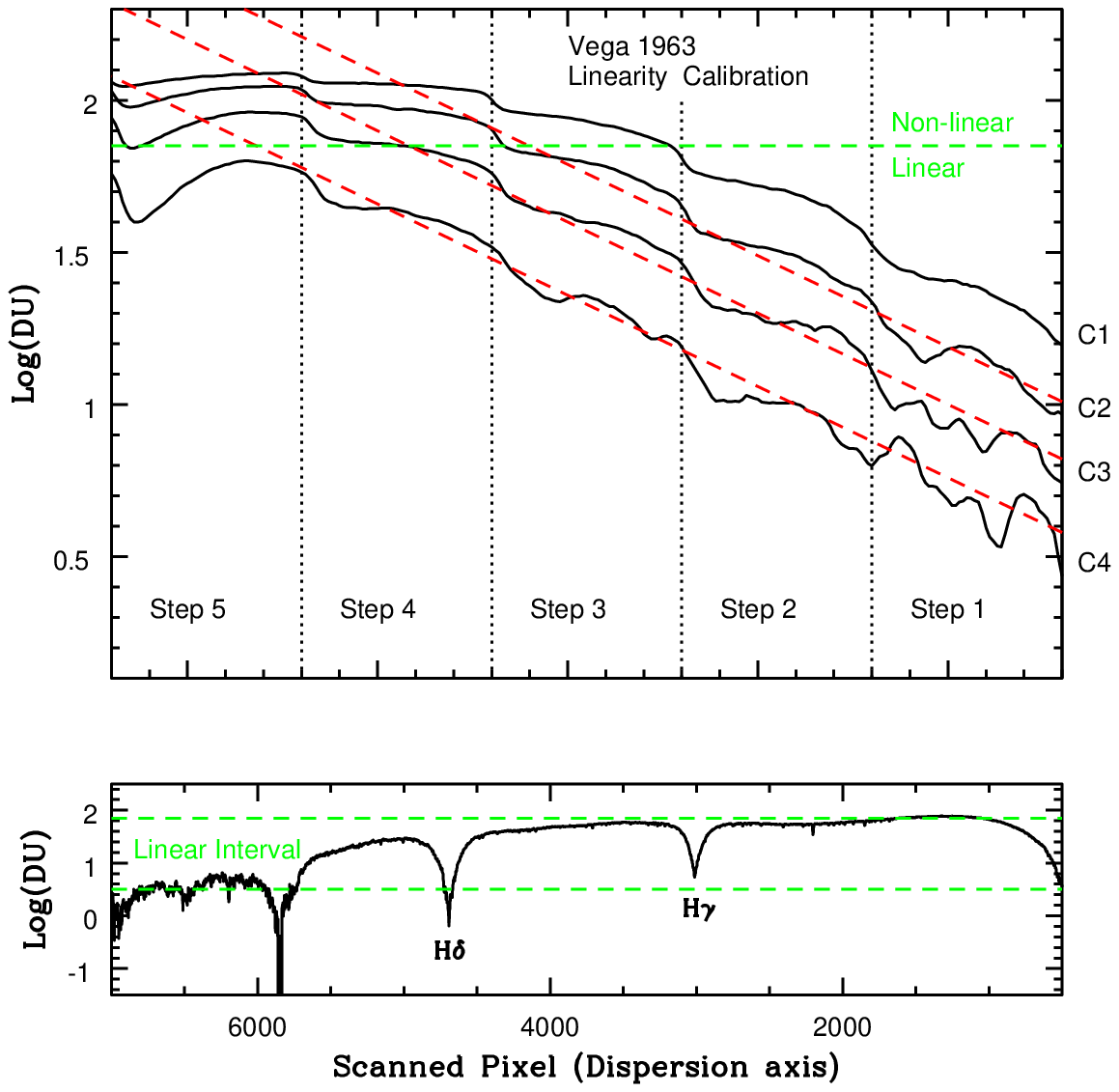}
\caption{(Top panel) Information extracted from the characteristic curve 
calibration region on the 1963 plate. Mean signal 
levels in four 10.5 mm wide strips, labeled C1 through C4, 
that cut through the calibration area are shown. 
The calibration region consists of five steps with signal levels that vary 
along the dispersion direction, and the boundaries between the steps 
in the region examined here are indicated by the 
vertical dotted lines; scattered light and the Eberhard effect blur the 
step edges and tilt the signal within each step. 
The signal in steps 4 and 5 is more-or-less flat 
in C1 and C2, suggesting that the signal is close to saturation and samples 
the upper part of the characteristic curve. Flattening at the bright end is 
much less noticeable in C3 and C4, which sample lower signal levels. 
A linear relation through the midpoints of each step with a slope 0.3 dex 
per step will be present if the signal is linear, and the 
dashed red lines are relations with a slope 0.3 dex/step that were fit to the 
central light levels in steps 1 -- 4; this relation was not fit to C1 because 
of the obvious nonlinearity at the bright end of this curve. The linear 
relations match the curves at signal levels log(DU) $\lessapprox 1.85$, 
indicated with the dashed green line, which we 
identify as the upper limit for a linear response. The comparisons 
in the top panel also indicate that the signal is 
linear down to light levels of at least log(DU) $\sim 0.5$.
(Bottom panel) The partially processed mean 1963 spectrum. 
The dashed green lines mark the interval over which the signal is expected 
to be linear, based on the comparisons in the top panel. The 
line at log(DU) $= 0.5$ is an upper limit. Note that H$\gamma$ 
and much of the continuum fall within the linear interval. 
Moderately weak absorption features in the high S/N part of the spectrum 
also fall within the linear boundaries, although they are close to 
the upper limit for linearity. The applicability of the linear 
response region is tested through comparisons with a CCD spectrum 
in Section 6.4.}
\end{figure}

	There is a factor of 2 (i.e. 0.3 dex) 
difference in signal between successive steps in the calibration 
region. Non-linearity is then evident if the difference in signal 
between successive steps departs from 0.3 dex. Evidence for 
nonlinearity at the bright end is clearly seen in Figure 10 as the 
signals in Steps 4 and 5 in C1 and C2 are similar. 
However, at lower signal levels C1 and C2 follow a relation that is closer 
to that expected for a linear response. 

	The red dashed lines in Figure 10 are linear fits to the signal 
measured in Steps 1 -- 4 with the slope fixed at 0.3 dex/step; 
this relation was not fit to C1 as nonlinearity is clearly an issue near the 
highest signal levels in that curve. C3 and C4 follow the 0.3 dex/step relation 
between steps 1 and 4. The match with C2 is poorer, 
as the signal in step 4 departs markedly from the trend defined 
by steps 1 -- 3. These comparisons suggest that 
the upper limit for linearity on this plate occurs near log(DU) 
$\sim 1.85$, and this threshold is indicated by the green dashed line. 
As for the lower limit to the linearity interval, the linearity 
curves can not be traced reliably to light levels log(DU) $< 0.5$. Hence, 
we identify log(DU) = 0.5 as a {\it preliminary} lower boundary for linearity.

	The portions of the 1963 spectrum that fall within the estimated 
boundaries for a linear response are indicated in the bottom panel of Figure 
10, where the mean 1963 spectrum from Figure 4 is shown with a logarithmic 
intensity scale. The four spectra on the 1963 plate have more-or-less similar 
signal levels, and so should have similar linearity boundaries. 
Much of the 1963 spectrum falls within the linear interval 
identified from the calibration information, with the upper limit 
for linearity falling close to the continuum redward of H$\gamma$. 
H$\gamma$ falls entirely within the expected linearity interval.

	The 1937 spectrum lacks calibration information for the 
characteristic curve, and this omission illustrates the incomplete nature of 
calibration information in the DAO collection. The absence of this 
calibration information is frustrating given that the Balmer lines in 
the 1937 spectrum have markedly different depths than those in 
the other spectra. \citet{ohm1940} and \citet{fur1973} describe a 
procedure to track response using information obtained directly 
from spectra. However, this procedure requires that the spectra 
have a constant dispersion along the slit, and this 
is not the case with the 1.8 meter spectra of Vega (Figure 3). An alternate 
means of correcting for nonlinearity might be to compare the photographic 
spectrum of a bright star that has a stable spectral energy distribution (like 
Vega) that was recorded on the same night as the science target with 
a CCD spectrum of that same star, processed to have the same wavelength 
resolution as the photographic spectrum. Such a comparison will yield a generic 
characteristic curve for the plates used on that night.

\subsection{Comparisons with a CCD Spectrum}

	Comparisons with spectra of Vega that were 
recorded with CCDs are an important means of assessing 
the scanned spectra. A spectrum of Vega that was observed with the McKellar 
spectrograph on the DAO 1.2 metre telescope was selected as a reference for 
such a comparison. The spectrum and associated calibration data were recorded 
on the night of September 6, 2013 as part of program 2013C10 (PI: Soydugan). 
This particular observation was selected because (1) the wavelength 
coverage is centered near H$\gamma$, and (2) the spectral resolution 
($\frac{\lambda}{\Delta\lambda} \sim 6000$) is comparable to that of 
the 1963 Vega spectrum. 

	The CCD spectrum was processed by following a standard sequence 
of steps that included bias subtraction, the suppression of cosmic rays, 
flat-fielding, the combination of signal in the image slicer elements, 
wavelength calibration, and normalization to a pseudo-continuum. 
The final processed CCD spectrum at wavelengths near H$\gamma$ 
is shown in Figure 11, where it is compared with a spectrum 
that is the average of the four spectra extracted from the 1963 plate. 
The wavelength resolutions of the CCD and photographic spectra are similar, 
but not identical.

\begin{figure}
\figurenum{11}
\epsscale{1.0}
\plotone{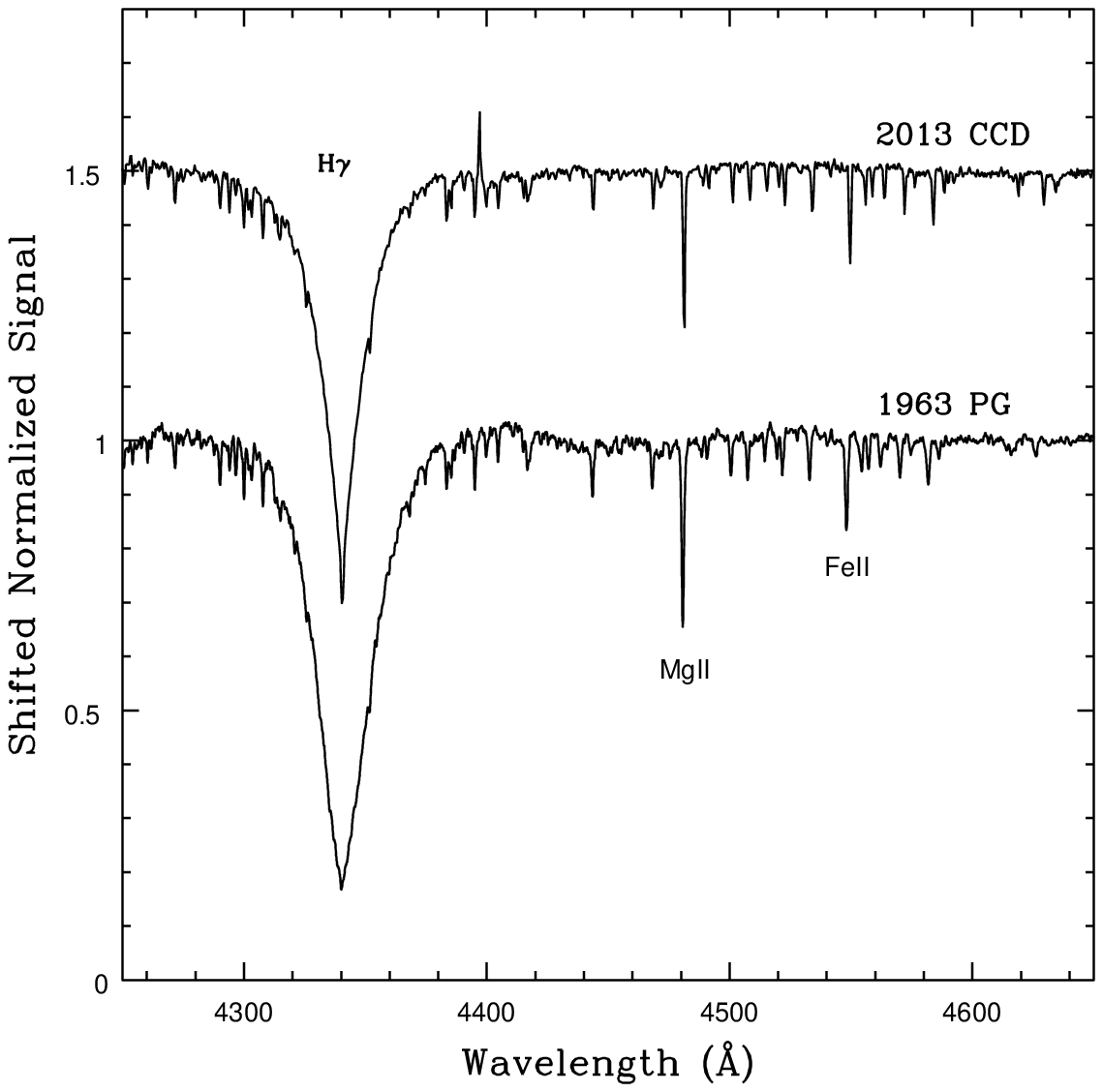}
\caption{Comparing a CCD spectrum of Vega taken with the DAO 1.2 
meter telescope with the mean spectrum extracted from the 1963 
photographic plate. The two spectra have similar (but not identical) 
wavelength resolutions, and the difference in line profiles is 
attributed in part to the illumination profiles of the spectra (see text).}
\end{figure}

	The H$\gamma$ line profiles in the 1963 and CCD spectra 
are markedly different, and we suspect that this is 
due in part to the variation in spectral resolution with location 
along the slit in the photographic spectrum, as shown in Figure 3. 
Combining spectra with different wavelength resolutions will produce 
an effective illumination profile for the photographic spectrum that differs 
systematically from that generated by an imager slicer-fed slit, which was 
used for the CCD observations. This will affect all lines in the 
photographic spectrum, and it is shown below that the lines in the 
photographic spectrum are systematically deeper than those in the CCD spectrum. 
While a more conventional line profile could be obtained for the 1963 spectrum 
by sampling only one location along the slit, the S/N is then 
greatly reduced. Another approach might be to broaden the spectra 
extracted from different slit locations to a common resolution. 
In any event, the difference in H$\gamma$ line profiles in 
Figure 11 is likely not due to the scanning process, but to properties that are 
intrinsic to the observations.

	There are numerous features in the CCD and photographic 
spectra that have small or modest depths, and these are featured in Figure 
12, where spectra between 4450 and 4600\AA\ are shown. There is an excellent 
one-to-one correspondence between weak features in the two spectra, 
many of which have depths that depart from the continuum by only a few percent. 
The depths of the weakest features in the photographic spectrum 
appear to match those in the CCD spectrum, although we caution that
differences between the depths of lines in the two spectra become 
progressively harder to detect by eye as line strength diminishes.

	The discussion of on-plate calibration information 
in Section 6.3 suggests that the features in Figures 11 and 12 fall in the 
range of intensities where a linear response is expected. 
In the absence of blending, the relative strengths of atomic lines is 
one means of comparing the science content of spectra that is less sensitive to 
differences in slit illumination. A visual inspection of 
Figure 12 reveals that the ranked strengths of various lines 
in the 1963 and CCD spectra are similar. That is, the 
strongest lines in the 1963 spectrum are also the strongest lines in the 
CCD spectrum, and a similar situation holds for the weakest lines. 

\begin{figure}
\figurenum{12}
\epsscale{1.0}
\plotone{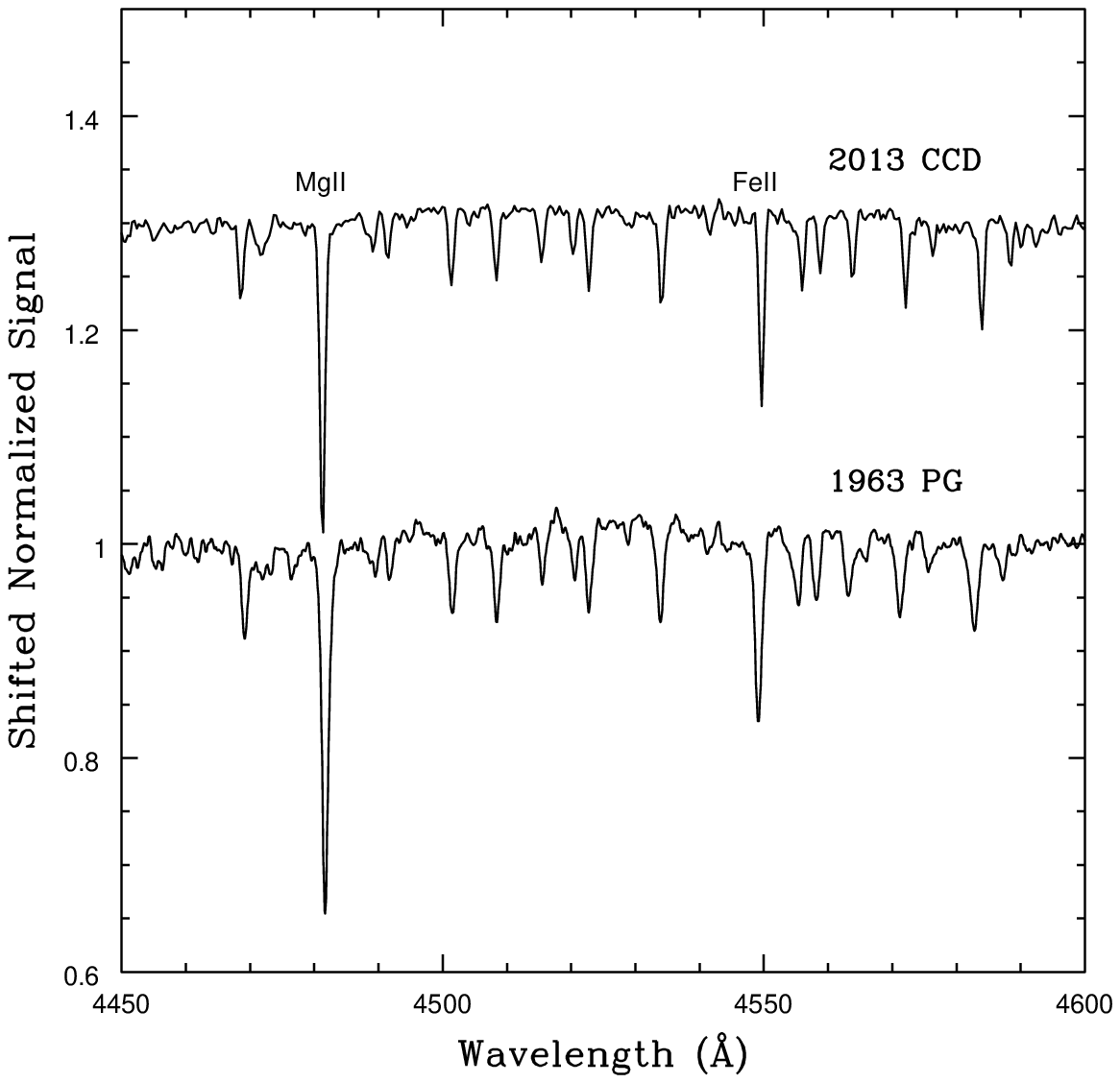}
\caption{Moderately weak absorption lines in the CCD and 1963 spectra. 
The scanned photographic spectrum contains weak features 
that match those in the CCD spectrum.}
\end{figure}

	If the response of the photographic spectrum is linear 
at the wavelengths shown in Figures 11 and 12 then the 
relative depths of lines should be the same in the two spectra. The 
equivalent widths of H$\gamma$, MgII 4481, and FeII 4542 in both spectra are 
listed in Table 2. These equivalent widths were measured with the SPLOT task 
in IRAF using the 'e' option, which computes equivalent widths
by integrating the absorbed light between user defined continuum points 
without assuming a line shape. This approach was adopted as efforts to fit 
lines in both spectra with Gaussian, Vogt, and Lorentzian line 
profiles were not satisfactory.

\begin{center}
\begin{deluxetable}{lccc}
\tablecaption{Comparing Equivalent Widths in the CCD and 1963 PG Spectra of Vega}
\tablehead{Spectrum & EW$_{H\gamma}$ \tablenotemark{a} & EW$_{MgII}$ \tablenotemark{b} & EW$_{FeII}$ \tablenotemark{b} \\
 & (\AA) & (\AA) & (\AA) }
\startdata
Vega PG & 24.17 & 0.44 & 0.22 \\
Vega CCD & 16.61 & 0.30 & 0.17 \\
 & & & \\
Ratio & 1.46 & 1.47 & 1.29 \\
 & $\pm 0.12$ & $\pm 0.18$ & $\pm 0.29$ \\
\enddata
\tablenotetext{a}{Estimated Uncertainty $\pm 0.5\AA$.}
\tablenotetext{b}{Estimated Uncertainty $\pm 0.03\AA$.}
\end{deluxetable}
\end{center}

	Uncertainties in the equivalent widths were estimated by varying 
the continuum points within a reasonable range based on the visual inspection 
of the spectra and then assessing the dispersion in the resulting equivalent 
widths. The similarity of the CCD and photographic spectra made it possible to 
place continuum points at similar locations in each spectrum, thereby 
reducing systematic errors that might arise if weak absorption features 
affect continuum placement in one spectrum but not the other.
These experiments suggest an uncertainty in both spectra of 
$\pm 0.5\AA$ for the H$\gamma$ measurements, and $\pm 0.03\AA$ for the MgII 
and FeII measurements. 

	The ratios of the two sets of equivalent widths are 
shown in the bottom row of Table 2. The ratios of all three lines 
agree to within their uncertainties, independent of line strength. This is 
consistent with the photographic spectrum at these wavelengths 
falling on the linear part of the characteristic curve, 
as predicted in Figure 10. The absolute difference between the 
strengths of individual features is reminiscent of what was seen in Section 
6.2, where spectra extracted from the same plate were compared. 
These differences aside, the relative consistency between the 
strengths of features in the photographic and CCD spectra 
in Figures 11 and 12 is noteworthy given the differences in the manner with 
which the spectra were recorded and processed. 

\section{DISCUSSION \& SUMMARY}

	The digitization of photographic spectra of Vega recorded with 
the 1.8 meter telescope in the DAO plate collection 
has been discussed, with the purpose of determining if 
commercial scanners can capture the scientific content of these 
spectra. The use of scanners to digitize photographic spectra 
opens the prospect of scanning historically important photographic 
material for scientific use in a straightforward, 
quick, and efficient manner. Investments in time, money, and 
space are modest with these devices. An additional benefit is ease of use once 
procedures and parameters such as scanning density have been established. 
Spectra were digitized with an Epson V800 device, which is a compact desktop 
scanner, and an Epson 12000XL, which is a physically larger top-of-the 
line scanner intended for artistic and commercial applications. 

	Spectra of Vega were selected for this study given (1) the relative 
stability of its spectral-energy distribution with time, which 
makes direct comparisons with CCD spectra possible, (2) 
the presence of absorption lines that span a range of intensities, and 
(3) the large number of high-quality spectra of this star in the 
plate collection. The plates digitized in the current study were recorded 
during the first four decades of the operation of the 1.8 meter 
telescope. During this time there were changes in photographic 
emulsions, hypersensitization techniques, and instrument optics. 
The spectra considered here thus have diverse properties, but are 
more-or-less representative of spectra that were recorded during the first 
half century of the operation of the facility. 

	Scattered light from the illuminating source in the scanner is a 
potential concern in the scanning process, as the grains in a photographic 
emulsion may scatter light. The possible impact of such scattering has been 
mitigated in this study through the subtraction of background light measured 
across the plate. Scattered light was removed in 
the processing. Another mitigation strategy is to extract 
spectra where the stellar signal is highest, although residual scattered light 
could still affect the cores of deep lines.

	A complicating factor for assessing the scientific utility of 
the spectra is that many plates in the DAO collection were recorded 
with the sole purpose of measuring radial velocities, with an 
emphasis on the measurement of line locations, rather than line shape. The need 
to measure radial velocities drove many of the design decisions for the 1.8 
meter telescope and the initial spectrograph configuration 
\citep[e.g.]{pla1918}, and these decisions have implications for 
extracting spectra for other purposes. For example, the emission 
line arcs are positioned close to the stellar 
spectra, presumably to facilitate reliable wavelength measurements 
with the technology available one hundred years ago. Scattered 
light from the arc overlaps the stellar spectra, and so will affect line 
depth measurements. Such contamination can be reduced (but not eliminated) 
by extracting spectra where the signal is highest. Complications such as 
this make it essential that 1.8 meter spectra that are scanned for archival 
purposes cover two dimensions.

	There are numerous other complications when extracting 
scientific information from photographic spectra. Arguably the greatest of 
these is the non-linear response characteristics of photographic 
emulsions. The majority of the plates in the DAO collection contain at least 
some calibration information that allows a characteristic 
curve to be constructed, although the sampling may be coarse, such as 
in the 1925 spectrum (e.g. Figure 1). While no attempt is made to 
apply a correction for nonlinearity in this study, the on-plate 
calibration information has been used to identify the range of 
linear signal levels, with the 1963 spectrum serving as an example.

	An important test of any digitization effort is consistency with CCD 
spectra, and the agreement between the 1963 spectrum digitized at 
2400 dpi with the 12000XL scanner and that recorded with 
a CCD is noteworthy. The scanned 1963 spectrum contains 
information not only about the deepest and widest features, which are the 
Balmer lines, but also of much weaker lines, such as those of MgII 4481 and 
FeII 4542. That the digitized 1963 spectrum contains information on the same 
spatial scale as the pixels in the CCD spectrum indicates 
that the information content on that DAO 1.8 meter plate has been recovered. 
It is encouraging that the ranked depths of weak lines in the CCD 
and photographic spectra match, although the equivalent widths differ. 
Still, the ratio of the equivalent widths of 
the FeII, MgII, and H$\gamma$ in the two spectra 
are similar, as expected if the response is linear in this 
part of the 1963 spectrum. This is consistent with measurements made from the 
linearity calibration region.

\subsection{Lessons Learned and Future Work}

	In closing, this paper has explored the use of commercial 
flatbed scanners to digitize photographic spectra in the DAO collection. These 
scanners offer a potential alternative to devices that have traditionally 
been used to digitize spectra, such as PDS machines. Here, we summarize 
some of the lessons that have been learned.

\begin{enumerate}

\item{There are commercially available flatbed scanners that 
can recover the scientific information contained in a stellar spectrum 
on a photographic plate.}

\item{A number of commercial flatbed scanners are available. A key 
characteristic when selecting a scanner to digitize plates is the 
MTF. We find that the MTF of the Epson 12000XL is superior to that 
of the Epson V800 at the scanning densities that are required to fully 
capture the scientific content on a 1.8 meter spectroscopic plate.}

\item{A scanning density of 2400 dpi recovers the information on 
plates recorded with the Cassegrain spectrograph on the DAO 1.8 meter 
telescope. This corresponds to a spatial resolution of $\sim 10\mu$m, 
which is similar to the grain size of most photographic 
emulsions that were in use throughout the 
first $\sim 50$ years that the telescope was in use. A higher 
scanning density might be appropriate for photographic plates with finer 
emulsions. However, we caution that scanning at too high a density might 
introduce digitization noise.}

\item{While not recommended for detailed scientific analysis, 
spectra scanned at lower spatial densities and with 8 bit sampling provide 
information that can be used to assess the scientific potential of a plate 
prior to more detailed analysis. This might prove useful for an initial 
assessment of information in large spectroscopic plate collections that have 
not yet been digitized.}

\item{Plates should be scanned in their entirety to recover the 
light profile of the target source as well as calibration information. 
The seemingly blank regions on a plate also provide information needed 
to understand background uniformity.}

\item{The linear part of the characteristic curve can be extracted 
from calibration information on the plate, although scattered light 
may blur the gradations. Large parts of this calibration region 
may also be saturated, so there is a need to be selective to determine 
where linearity information is to be extracted. The use of such on-plate 
information is an important check of possible plate-to-plate differences in the 
characteristic curve.} 

\item{Although recorded with the same instrumental configuration 
and photographic emulsion, the depth of H$\gamma$ in the 
1925 and 1937 spectra are very different. More subtle, but still significant, 
differences are also seen when comparing spectra that were recorded on the same 
plate. This demonstrates that spectra of the same source in the DAO 
collection may differ for reasons that are not intrinsic to the source.}

\item{While promising, the comparisons in Figures 11 and 12 indicate that 
additional processing is required to match line strengths with 
those in CCD spectra. This highlights that there are other issues that must be 
addressed once plates are digitized. In addition to dealing with the inherent 
problems associated with the use of photographic 
materials, there is also the need to capture meta data 
and develop tools to facilitate the scientific 
analysis of photographic spectra.}

\end{enumerate}

	There is much room for future work. The large content of the DAO plate 
library, of which spectra obtained with the 1.8 meter telescope is but one 
part, presents a formidable logistical challenge for digitization efforts, 
and strategies for digitizing the collection in 
an effective manner must be explored. In addition to 
spectra, there is also a collection of photographic images in the 
DAO collection that were recorded at the Newtonian focus 
of the 1.8 metre telescope over a number of decades, and these 
also contain a wealth of information. As demonstrated by \citet{gluetal2022} 
and \citet{ceretal2021}, the digitization techniques discussed 
here could be applied with advantage to those images. 

\acknowledgements{It is a pleasure to thank the anonymous referee for 
providing comments that greatly improved the manuscript.}

\end{document}